\documentclass[%
 aip,
 amsmath,amssymb,
 reprint,%
]{revtex4-1}

\usepackage{graphicx}
\usepackage{dcolumn}
\usepackage{bm}

\usepackage[utf8]{inputenc}
\usepackage[T1]{fontenc}
\usepackage{mathptmx}

\begin{document}

\preprint{AIP/123-QED}

\title[An improved equation of state for air plasma simulations]{An improved equation of state for air plasma simulations}

\author{F. Tr{\"a}uble}
 \altaffiliation[Also at ]{Max Planck Institute for Intelligent
   Systems, T{\"u}bingen, Germany}
\author{S. T. Millmore}%
\email{stm31@cam.ac.uk}
\author{N. Nikiforakis}
\affiliation{ 
Cavendish Laboratory, University of Cambridge, Cambridge, United Kingdom
}%

\date{\today}

\begin{abstract}
  This work is concerned with the development of a novel, accurate
  equation of state for describing partially ionised air plasma in
  local thermodynamic equilibrium.  One key application for this new
  equation of state is the simulation of lightning strike on aircraft.
  Due to the complexities of species ionisation and interaction,
  although phenomenological curve fitting of thermodynamic properties
  is possible, these curves are intractable for practical numerical
  simulation.  The large difference in size of the parameters (many
  orders of magnitude) and complexity of the equations means they are
  not straightforward to invert for conversion between thermodynamic
  variables.  The approach of this paper is to take an accurate
  19-species phenomenological model, and use this to generate a
  tabulated data set.  Coupled with a suitable interpolation procedure
  this offers an accurate and computationally efficient technique for
  simulating partially ionised air plasma.  The equation of state is
  implemented within a multiphysics methodology which can solve for
  two-way coupling between a plasma arc and an elastoplastic material
  substrate.  The implementation is validated against experimental
  results, both for a single material plasma, and an arc coupled to a
  substrate.  It is demonstrated that accurate, oscillation-free
  thermodynamic profiles can be obtained, with good results even close
  to material surfaces.
\end{abstract}

\maketitle

\section{Introduction}

Numerical simulation of air plasma interacting with solid substrates
provides insight into the physical mechanisms which lead to structural
damage due to lightning strike, which is a key safety concern for
aircraft~\cite{morgan2012interaction}.  The aluminium skin
traditionally used for aircraft rapidly dissipates the energy input
from the lightning strike, due to its high electrical and thermal
conductivity.  With the current trends towards lightweight composite
skins, with lower conductivity, there is greater energy input local to
the lightning attachment point.  To accurately understand the physical
effects at this attachment, an accurate description of the ionisation
processes within the plasma arc is essential.  This requires an
equation of state (EoS) which can account for the complex
thermodynamic and electromagnetic effects arising from partially
ionised material.  The multiple atomic species comprising an air
plasma are each governed by different characteristic ionisation
energies, governing their initial emergence and further ionisation.
This leads to a non-linear relationship between intensive thermodynamic
variables within the EoS for an air plasma.  Computation of these
quantities, which on a continuum scale model of a plasma are local
physical properties, and are independent of system size, must still
take into account transport properties arising from molecular
interactions between the different species present.

It is a key challenge in developing an EoS for air plasma to balance
accuracy with computational efficiency.  One of the simplest
techniques is to use a surrogate ideal gas model with an adiabatic
index appropriate for much of the ionised regime, for example Ekici et
al.\ \cite{ekici2007thermal} use a an effective adiabatic exponent of
$\gamma=1.16$.  In a lightning arc simulation, for which there is the
transition of unionised air to a plasma, an ideal gas model has
limited validity.  An improved model was developed by Plooster
\cite{plooster1970shock} and has been applied in several plasma
discharge models. It is a simplified equation of state for a surrogate
diatomic gas, which takes into account dissociation effects. However,
due to the different dissociation energies for $N_{2}$ and $O_{2}$,
key components of an air plasma, and the fact that nitric oxide ($NO$)
is not considered, this approach is limited to lower temperature
mixtures, below $9,000\text{K}$ \cite{akram1996evolution}. Villa et
al.\ \cite{villa2011multiscale} developed an EoS which models 11 of
the most common species within an air plasma, allowing for single
ionisation of $N_{2}$ and $O_{2}$, as well as the dissociated atoms.
Each species is treated as an ideal gas which are then combined as a
generalised ideal gas mixture.  They start from the work of Mottura et
al.\ \cite{mottura1997evaluation} who assume a simplified heat
capacity of the form $C_{v}(T)$ with $T$ being the temperature,
neglecting effects from a variation in pressure. The equation of state
for each species assumes a linear relation between the specific heat
contribution and species' dependent formation heats. To fully describe
a given set of thermodynamic data, a system of coupled linear
equations must be solved.  In order to maintain computational
efficiency, Villa et al.\ tabulated the thermodynamic data for given
pairs of density and specific internal energy, and interpolate values
between these points.  This tabulated form allows for a complex
description of the plasma composition, though, in common with other
equations of state available in the open literature, it does not
consider plasma-specific effects arising from electromagnetic
interaction on microscopic scales.

In this work, an improved equation of state is developed based on an
accurate air plasma model covering a wide range of variables from the
theoretical model of D'Angola et al.\
\cite{d2008thermodynamic,d2011thermodynamic}.  In section 2 the
thermodynamic theory of air plasma, upon which the equation of state
is based, is detailed. Additionally, a mathematical formulation for
describing the governing equations of motion for an air plasma, which
will be used to validate the improved EoS, are given, using the model
developed by Millmore and
Nikiforakis~\cite{millmore2019multiphysics}. The generation of the new
EoS is detailed in section 3 and section 4 presents validation of the
improved EoS within a state-of-the-art multiphysics multimaterial
lightning code.  It is demonstrated that an accurate EoS is required
to model the arc attachment point to a substrate, and thus is
essential in order to gain a deeper insight into this highly complex
and nonlinear phenomenon. Conclusions and further work are presented in
section 5.

\section{Theory and mathematical formulation}

\subsection{Air plasma model\label{subsec:Air-plasma-model}}

Atmospheric air is a mixture of many different particle species
which, over the temperature and pressure range experienced in a
lightning strike, will interact in a complex manner
\cite{capitelli2013fundamental}. The composition of dry air per volume
is primarily made up of $78.084\%$ nitrogen, $20.946\%$ oxygen,
$0.934\%$ Argon and $0.036\%$ carbon dioxide. Additional species such
as other noble gases occur not more than a few parts per million
particles. Under realistic conditions, there is a variable amount of
water vapour present in the air
\cite{brimblecombe1996air,atmosphere1955data}.

An appropriately chosen theoretical model of air plasma has to cover the
physics between the limits of unionised low-temperature air and highly
ionised air plasma. The generation of the EoS presented here is based
on an accurate theoretical model for a 19-species air plasma model
developed by Capitelli, Colonna, D'Angola and others in
\cite{d2008thermodynamic,d2011thermodynamic,capitelli2011fundamental,capitelli2013fundamental,colonna2004hierarchical,colonna2007improvements}.
This air plasma model considers the following 19 air species:

\begin{equation*}
N_{2},N_{2}^{+},N,N^{+},N^{++},N^{+++},N^{++++}  
\end{equation*}
\begin{equation*}
O_{2},O_{2}^{+},O_{2}^{-},O,O^{-},O^{+},O^{++},O^{+++},O^{++++}  
\end{equation*}
\begin{equation*}
NO,NO^{+},e^{-}  
\end{equation*}
Molecular species with more than two atoms are not considered, as they
only occur in low quantities, so their effect on the application
presented here is negligible, and the theory for excited energy levels
of these species is complex. The chemical processes among the
different species are assumed to be in detailed balance at all times
with the only exception being radiation processes, which involve the
emission and absorption of photons. This assumption is valid as long
as molecular timescales are significantly smaller than fluid and gas
flow timescales; this is satisfied under the thermodynamic conditions
of interest. In this case, a unique temperature and pressure can be
assigned to a given composition. As such intensive variables are
inhomogeneous in time and space for the applications of interest in
this paper, the concept of local thermodynamic equilibrium (LTE) needs
to be defined. LTE requires that the particles' energies in a local
neighbourhood obey Maxwell-Boltzmann distributions, so that a local
temperature and pressure can be defined. This is not always perfectly
justified, but work from Haidar \cite{haidar1999non} indicates that it
is sufficiently accurate to predict the expected behaviour in a
lightning attachment.

\subsubsection{Thermodynamics\label{subsec:Thermodynamics}}

In dynamical equilibrium, every reaction is governed by a law of mass
action, taking the form
\begin{equation}
K_{r}^{P}(T)=\prod_{s=1}^{N_{\text{total}}}P_{s}^{\nu_{r,s}}\qquad r=1,2,...,N_{R}\label{eq:equilibrium_equations}
\end{equation}
where $K_{r}^{P}(T)$ is a temperature dependent equilibrium constant
of the $r^{th}$ reaction and $P_{s}$ is the partial pressure of
the $s^{th}$ species. $N_{\text{total}}$ is the number of considered
species and $\nu_{r,s}$ the stoichiometric coefficients of the $s^{th}$
species in the $r^{th}$ reaction. The total pressure of a gas mixture
is then given by Dalton's law 
\begin{equation}
P=\sum_{s=1}^{N_{\text{total}}}P_{s}+P_{DH}\label{eq:total_pressure_equation}
\end{equation}
where the correction term $P_{DH}$, derived from the Debye-Hückel
theory, considers additional effects for ionised gas mixtures, with
further details below. Every species is partially approximated by
$P_{s}=n_{s}k_{B}T$ with $n_{s}$ the particle number density of
the $s^{th}$ species, $k_{B}$ the Boltzmann constant and $T$ the
temperature. The particles are modelled as dimensionless hard spheres
and virial corrections are neglected. This is justified as they are
usually only important for $T<300\text{K}$ with pressure of $P\approx100\text{atm}$
\cite{hilsenrath1965tables,capitelli1977thermodynamic,sevast1986virial},
i.e.\ unrealistic conditions for the arc channel plasma and surrounding
air.

For given pressure and temperature, the system of nonlinear equations
arising from \eqref{eq:equilibrium_equations} and \eqref{eq:total_pressure_equation}
determines the exact composition of all the species present in the
air model. A further condition is given by the conservation of the
predetermined total mass of each atomic species, adding two additional
constraints to the 17 equations to close the system. It is challenging
to solve such a nonlinear system numerically, since equilibrium constants
differ by many orders of magnitude. The composition used for the numerical
fits in \cite{d2008thermodynamic,d2011thermodynamic} is based on
an accurate hierarchical algorithm for solving the equilibrium of
a given reactive mixture \cite{d2011thermodynamic}. 

Once the composition $\{n_{s}\}_{s=1,...,N_{total}}$ is known, the
thermodynamic potentials are calculated using the respective partition
function for each species
\cite{capitelli2011fundamental,d2008thermodynamic}.  Due to the
formation of a plasma potential in ionised gases, and plasma in
general, the thermodynamic properties are modified when compared to an
ideal gas. The theory and framework to consider such effects were
developed in the Debye-Hückel theory \cite{landau1959statistical}.
This introduces a cut-off energy, making the sum concerning the
internal partition functions finite and further correction terms for
the pressure and potential\cite{capitelli2011fundamental}. Finally,
the heat capacities are given by
\begin{align}
C_{P} & =\Bigg(\frac{\partial H}{\partial T}\Bigg)_{P}\label{eq:cp}\\
C_{V} & =\Bigg(\frac{\partial U}{\partial T}\Bigg)_{V}\label{eq:cv}
\end{align}
with $H$ being the enthalpy and $U$ the internal energy.

\subsubsection{Transport properties\label{subsec:Transport-properties}}

In the presence of gradients in the spatial distribution of thermodynamic
quantities, transport effects emerge which act to reduce these gradients.
Each thermodynamic behaviour has a different responsible transport
mechanism. Ordinary diffusion is associated with a concentration difference
and thermal diffusion connected to a gradient in temperature. Viscosity
is the mechanism describing the transfer of momentum due to velocity
gradients, while thermal conductivity contains effects that lead to
transport in thermal energy such as thermal gradients, chemical reactions
or redistributions of internal degrees of freedom. Electrical conductivity
is related to the transport of charged particles due to differences
in the electric potential. At energies relevant to air plasmas within
this work, it is justified to assume that the transport coefficients
of electrons and heavy atoms or ions decouple from each other \cite{capitelli2000transport}.
Numerical fits regarding the transport coefficients in \cite{d2008thermodynamic,d2011thermodynamic}
have been calculated using third-order approximated coefficients derived
from the Chapman-Enskog theory \cite{d2008thermodynamic,devoto1966transport}.
The Chapman-Enskog theory provides a set of equations for calculating
the transport properties of a multi-species gas mixture under the
assumption that thermodynamic states can be described in LTE. Its
starting point is the Boltzmann equation formulation and it assumes
some justified microscopic models for the binary collision term \cite{chapman1970mathematical}.
The main approach of the theory represents a Chapman-Enskog expansion
where the probability density function is expanded perturbatively
in a small parameter $\varepsilon$. This will result in a general
set of Navier-Stokes equations for this system, including terms for
transport coefficients. A detailed derivation can be found in
Capitelli et al.\cite{capitelli2013fundamental},
and this derivation provides expressions for characteristic transport
coefficients to an arbitrary degree of order $\xi$. In this work,
the transport coefficients of interest are electrical and thermal
conductivity\cite{capitelli2000transport}. Corresponding collision
integrals, as calculated by D'Angola et al., take into account a variety
of species-depended interaction approximations.

\subsection{Mathematical formulation of a plasma discharge\label{subsec:Mathematical-formulation-plasma-discharge}}

\begin{figure}[tbh]
\begin{centering}
\includegraphics[width=0.9\columnwidth]{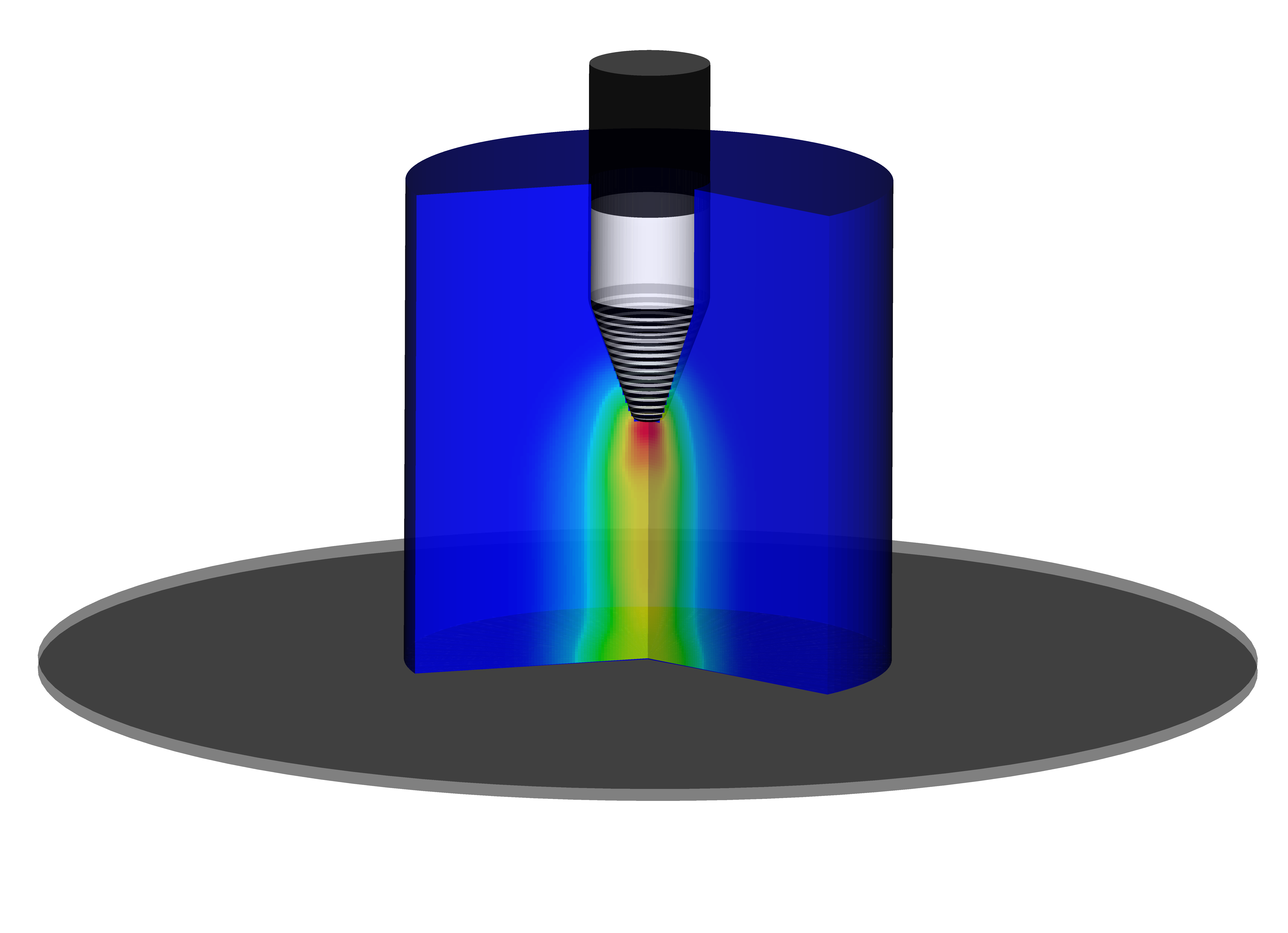}
\par\end{centering}
\caption{Schematic showing the axisymmetric plasma discharge setup. A
  plasma arc is generated at the electrode at the top of the image,
  and connects to the substrate beneath it.  A cut-through of the
  air/plasma region is shown, illustrating the arc; the plasma does
  occupy the full domain, to the edge of the substrate.  Beneath the
  substrate is a subsequent air region (not shown), an approximate
  two-dimensional representation of this setup is shown in
  Figure~\ref{fig:Geometry__2D_plane}. \label{fig:IllustrationPlasmaDischarge}}
\end{figure}

To model the dynamics of a plasma discharge for lightning studies, a
magnetohydrodynamic formulation is used.  This assumes the plasma is
modelled as a single fluid with a corresponding equation of state that
considers thermodynamics of interacting species; this is appropriate
for an air mixture in LTE.  The governing equations are
\begin{align}
\frac{\partial}{\partial t}\rho+\nabla\cdot(\rho\bm{u}) & =0\label{eq:mass balance equation}\\
\frac{\partial}{\partial t}(\rho\bm{u})+\nabla\cdot(\rho\bm{u}\otimes\bm{u}+P\bm{I}) & =\bm{J}\times\bm{B}\label{eq:force balance equation}\\
\frac{\partial}{\partial t}E+\nabla\cdot(\bm{u}(E+P)) & =\bm{u}\cdot(\bm{J}\times\bm{B})+\rho_{res}|\bm{J}|^{2}-S_{r}\label{eq:energy balance equation}\\
\nabla^{2}\bm{A} & =-\mu_{0}\bm{J}\label{eq:poisson equation}
\end{align}
with density, $\rho$, velocity ${\bm u}$, total energy $E=\rho
e+\frac{1}{2}\rho\bm{u}^{2}$, magnetic potential, ${\bm A}$, magnetic
field $\bm{B}=\nabla\times\bm{A}$, current density, ${\bm J}$, electrical resistivity $\rho_{res}=1/\sigma$,
electrical conductivity $\sigma$, permeability of free space
$\mu_{0}$, and a source term governing radiative losses, $S_r$.
As the system is underdetermined (9 unknowns in 8 equations), one
further equation is needed for closure. This is provided by an equation
of state for air plasma, providing the mathematical relation between
specific internal energy $(e)$, pressure $(P)$ and mass density,
in the form $e=e(\rho,P)$. Within this framework, a cylindrically
axisymmetric geometry is considered, with the arc connection at the
centre of the domain, which is sufficient to capture the bulk behaviour
of the arc-substrate interaction \cite{martins2016etude}. The system,
which is representative of a typical set up in lightning-protection
laboratory experiments, is schematically shown in Figure \ref{fig:IllustrationPlasmaDischarge}.
 Current profiles representative of those used in laboratory lightning
testing are included within the model through the current density.
Two different models are studied: a simplified 1D model and a more
realistic axisymmetric model.

Within the 1D model, radial arc profiles are reproduced by assuming
translational invariance along the z-coordinate. Consequently, the
current density has a component only in the $z$-direction and is
approximated by a Gaussian profile as suggested in \cite{villa2011multiscale}

\begin{equation}
\bm{J}(r,t)=-\frac{I(t)}{\pi r_{0}^{2}}e^{-(r/r_{0})^{2}}\bm{e}_{z}\label{external current density}
\end{equation}
where $r_{0}$ is the characteristic length scale of the plasma arc
channel radius and the total current $I(t)$ is a given input function.
By assuming such a predetermined profile, the resulting magnetic field
is not coupled to the fluid evolution itself. 

\begin{figure}[tbh]
\hfill{}%
\begin{minipage}[t][1\totalheight][b]{0.8\columnwidth}%
\includegraphics[width=1\columnwidth]{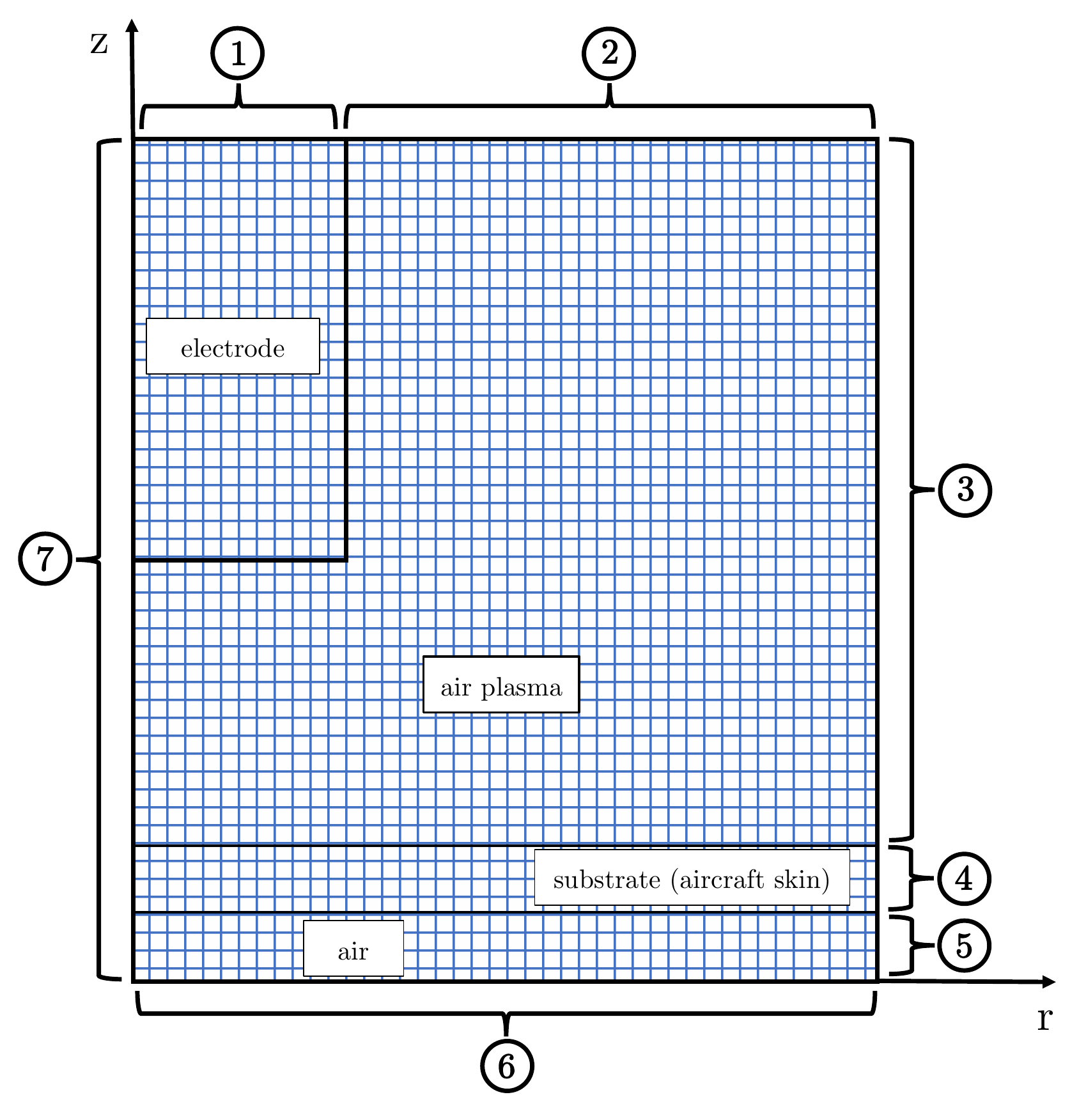}%
\end{minipage}\hfill{}

\hfill{}%
\noindent\begin{minipage}[t]{1\columnwidth}%
\begin{tabular}{cccc}
\toprule 
 & {\small{}$\phi$} & {\small{}$A_{r}$} & {\small{}$A_{z}$}\tabularnewline
\hline
\hline 
{\small{}1} & {\small{}$\frac{\partial\phi}{\partial n}=-\frac{1}{\sigma}\frac{I(t)}{\pi r_{0}^{2}}$} & {\small{}$A_{r}=0$} & {\small{}$\frac{\partial A_{z}}{\partial n}=0$}\tabularnewline
\hline 
{\small{}2} & {\small{}$\frac{\partial\phi}{\partial n}=0$} & {\small{}$A_{r}=0$} & {\small{}$\frac{\partial A_{z}}{\partial n}=0$}\tabularnewline
\hline 
{\small{}3} & {\small{}$\frac{\partial\phi}{\partial n}=0$} & {\small{}$\frac{\partial A_{r}}{\partial n}=0$} & {\small{}$A_{z}=0$}\tabularnewline
\hline 
{\small{}4} & {\small{}$\phi=0$} & {\small{}$\frac{\partial A_{r}}{\partial n}=0$} & {\small{}$A_{z}=0$}\tabularnewline
\hline 
{\small{}5} & {\small{}$\frac{\partial\phi}{\partial n}=0$} & {\small{}$\frac{\partial A_{r}}{\partial n}=0$} & {\small{}$A_{z}=0$}\tabularnewline
\hline 
{\small{}6} & {\small{}$\frac{\partial\phi}{\partial n}=0$} & {\small{}$A_{r}=0$} & {\small{}$\frac{\partial A_{z}}{\partial n}=0$}\tabularnewline
\hline 
{\small{}7} & {\small{}$\frac{\partial\phi}{\partial n}=0$} & {\small{}$\frac{\partial A_{r}}{\partial n}=0$} & {\small{}$\frac{\partial A_{z}}{\partial n}=0$}\tabularnewline
\hline
\end{tabular}
\end{minipage}\hfill{}

\caption{Geometry of the computational domain in the 2D model (left) including
boundary conditions for the electromagnetic potentials (right). \label{fig:Geometry__2D_plane}}
\end{figure}

Within the axisymmetric model, the geometry of an electrode is explicitly
included and the current density is computed dynamically. Hence, this
approach is more accurate as it calculates the magnetic field in a
plasma arc self-consistently since it is coupled with the evolution
of the air plasma. It also allows for interaction with the boundaries
to affect the current profile. The current density is connected to
the electric field $\bm{E}$ or its corresponding potential $\phi$
through
\begin{equation}
\bm{J}=\sigma\bm{E}=-\sigma\nabla\phi\label{eq:currentPotentialConnection}
\end{equation}
and due to the conservation of charge, it holds $\nabla\cdot\bm{J}=0$.
As a consequence, elliptic PDEs for $\phi$ and $\bm{A}$ have to
be solved, given by
\begin{align}
\nabla\cdot\big(\sigma\nabla\phi\big) & =0\\
\nabla^{2}\bm{A} & =\mu_{0}\sigma\nabla\phi
\end{align}
which represent some form of generalised Poisson's equations on the
domain. Figure \ref{fig:Geometry__2D_plane}
shows the domain and its geometry for the 2D axisymmetric model.
Applied boundary conditions for the electromagnetic potentials are
summarised in the table beneath the diagram. If an aluminium panel is
used as the aeronautical substrate, an electrical conductivity of
$\sigma=3.2\times10^{7}\text{Sm}^{-1}$ is used, i.e.\ higher than
typical air plasma conductivity within the arc channel. Alternatively,
this substrate will be modelled as a carbon fibre composite
material. A complete implementation of an anisotropic equation of
state for carbon fibre composites is beyond the scope of this
work. Instead, an approximation to CFRP is used,
following~\cite{millmore2019multiphysics}, which defines the material
as a isotropic material, which can be employed to look at behaviour
either perpendicular or parallel to the CFRP weave.  This material
differs from aluminium in electrical conductivity
($\sigma=1.6\times10^{4}\text{Sm}^{-1}$) \cite{tholin2015numerical}
and mass density ($\rho=1538\text{kg}/\text{m}^{3}$)
\cite{martins2016etude}.

The model described above is implemented within a more extensive
simulation code. This includes explicit thermomechanical coupling with
various aircraft layer configurations in a full multimaterial model
\cite{millmore2019multiphysics}. This uses the Riemann Ghost Fluid
Method (GFM) to accurately model the multimaterial boundary
conditions, i.e.\ propagation and reflection of shock waves across a
multimaterial interface. The location of the interface is tracked by a
level set function. The original GFM was developed by Fedkiw et al.\
in \cite{fedkiw1999non}, and a detailed introduction to Riemann GFM is
given in the work of Sambasivan et al.\ \cite{sambasivan2009ghost}.
Different materials are evolved independently from each other on their
respective domains, and a unique equation of state can be assigned to
each material. In this code, the elastoplastic substrate and electrode
are described using the Eulerian framework as presented by Schoch et
al. \cite{schoch2013eulerian} and Michael et
al. \cite{michael2018multi}, based on the formulation of Godunov and
Romenskii \cite{godunov1972nonstationary}.  Plasticity effects are
incorporated following the work of Miller and Collela
\cite{miller2001high}. Details are discussed in Millmore and
Nikiforakis \cite{millmore2019multiphysics}, which validate the
equations from the previous section.

\section{Methodology}

\subsection{Generating data for an improved equation of state}

D'Angola et al. \cite{d2008thermodynamic,d2011thermodynamic}
calculated a wide range of thermodynamic and transport quantities for
given temperature and pressure (mixture composition, molar mass,
enthalpy, heat capacities, electrical and thermal conductivity), based
on the air plasma model described in section
\ref{subsec:Air-plasma-model}. Therewith an improved equation of state
for a 19-species air plasma model was developed. The equation of state
is inserted into the plasma discharge model from section
\ref{subsec:Mathematical-formulation-plasma-discharge} by the help of
a tabular database, denoted by EOS19. In principle, EOS19 can be used
by any simulation code that attempts to model air plasma applications
in the future. Every data point in the database generated contains the
following quantities:
\begin{itemize}
\item mass density $\rho_{i}$
\item pressure $P_{i}$
\item specific internal energy $e_{i}$
\item temperature $T_{i}$
\item speed of sound $c_{i}$
\item adiabatic index $\gamma_{i}$
\item molar fractions of each species $\text{\ensuremath{\chi_{i}}}$
\item electrical conductivity $\sigma_{i}$
\item thermal conductivity $\kappa_{i}$
\end{itemize}
The actual relation for the equation of state is then given through
discrete tuples of the form $(\rho_{i},P_{i},e_{i})$ with $e(\text{\ensuremath{\rho_{i},P_{i})=e_{i}}}$.
To generate this data the remaining quantities not provided as analytical
fits in \cite{d2008thermodynamic,d2011thermodynamic} need to be calculated.
The mass density of air plasma $\rho$ is calculated with the help of
the mean molar mass $\bar{M}$ via the formula

\begin{equation}
\rho=\frac{P}{RT}\bar{M}\label{eq:density-from-mean-molar-mass}
\end{equation}
with $R$ being the gas constant. Specific internal energy is derived
from the specific enthalpy with the help of a Legendre transformation
according to
\begin{align}
e & =h-\frac{P}{\rho}\label{eq:intEne-enthalpy_relation}
\end{align}
Although negligible in this model, thermal conductivity from analytical
fits in \cite{d2008thermodynamic,d2011thermodynamic} is included
in the database, such that it can be used to considering associated
effects in the future. Finally, the adiabatic index and corresponding
speed of sound, $a$, for air plasma as suggested in \cite{colonna2012electronic,capitelli2011fundamental}
are calculated as

\begin{align}
\gamma & =\frac{C_{P}}{C_{V}}\frac{\rho}{P}\Big(\frac{\partial P}{\partial\rho}\Big)_{T}\label{eq:soundspeed_isentropic_index}\\
a & =\sqrt{\gamma\frac{P}{\rho}}\label{eq:soundspeed-equation}
\end{align}
with the heat capacities defined in equations \eqref{eq:cp} and \eqref{eq:cv}. 

For typical choices of conserved variables, it is desirable to generate
the thermodynamic EoS quantities for given pairs of density and pressure,
instead of temperature and pressure. Therefore, the temperature has
to be obtained inversely for given density and pressure in the form
$T(\rho,P)$ with the help of the known relation in \eqref{eq:density-from-mean-molar-mass}
\begin{equation}
\rho(P_{i},T_{i})=\rho_{i}\label{eq:DenPreTempRelation}
\end{equation}
where the tuple $(T_{i},P_{i},\rho_{i})$ represents a thermodynamical
state. Finding the solution for $T_{i}$ in \eqref{eq:DenPreTempRelation}
reduces to a root-finding problem, and a simple bisection method is
applied. It is not important to have an efficient algorithm here since
the tabulated data is generated only once. However, even with a more
advanced method, this root-finding process is time-consuming, which
is the reason for creating a tabulated database instead of calculating
parameters dynamically during simulation. From this data, the equation
of state is interpolated. The layout of EOS19 follows the SESAME Database
standard for equations of state as introduced at Los Alamos National
Lab \cite{sesame}. It contains data points ranging from $0.001\text{kg/m}^{3}$
to $10\text{kg/m}^{3}$ in mass density and $0.01\text{atm}$ to $180\text{atm}$
in pressure. Interpolation from this database was achieved by a binary
search algorithm. 

\subsection{Numerical approach for solving the plasma discharge}

For each computational time-step $\Delta t$, the system of equations
\eqref{eq:mass balance equation} - \eqref{eq:poisson equation} is
evolved in four steps, following the approach in
\cite{villa2011multiscale}.
\begin{itemize}
\item [{Step\ 1}] The Poisson's equation in \eqref{eq:poisson
    equation} for the magnetic vector potential $\bm{A}$ is solved
  using the Thomas-Algorithm in 1D and the finite element solver
  framework FEniCS \cite{alnaes2015fenics} in axisymmetry. Both
  algorithms are an appropriate choice regarding accuracy and
  efficiency, however any suitable solver could be used. In
  axisymmetry, the magnetic vector potential has the non-zero
  components $A_{z}(r,z)$ and $A_{r}(r,z)$. The corresponding magnetic
  field, pointing in the poloidal direction, is then given by
\begin{equation}
B_{\phi}=\big(\nabla\times\bm{A}\big)_{\phi}=\frac{\partial A_{r}}{\partial z}-\frac{\partial A_{z}}{\partial r}\label{eq:MagFieldDefinition}
\end{equation}
with $A_{r}=0$ in the 1D model.
\item [{Step\ 2}] The homogeneous part of the Euler problem in \eqref{eq:mass balance equation}
- \eqref{eq:energy balance equation}, given by 
\begin{align}
\frac{\partial}{\partial t}\rho+\nabla\cdot(\rho\bm{u}) & =0\label{eq:HomEq1}\\
\frac{\partial}{\partial t}(\rho\bm{u})+\nabla\cdot(\rho\bm{u}\otimes\bm{u}+P\bm{I}) & =0\label{eq:HomEq2}\\
\frac{\partial}{\partial t}E+\nabla\cdot(\bm{u}(E+P)) & =0\label{eq:HomEq3}
\end{align}
is solved using high-order shock-caputuring schemes.  In this work,
the second-order finite volume method SLIC (Slope LImiter Centred) is used \cite{toro2013riemann}.
\item [{Step\ 3}] To incorporate Joule heating and radiation source terms
in the energy balance equation \eqref{eq:energy balance equation},
an ordinary differential equation of the form 
\begin{equation}
\frac{dE}{dt}=\rho_{res}|\bm{J}|^{2}-S_{r}\label{eq:Step3-Eq1}
\end{equation}
has to be solved, with details regarding the implementation of the
radiation model given in \cite{villa2011multiscale,VillaSourceCode}. 
\item [{Step\ 4}] To include Lorentz force effects, the ODEs given by
\begin{align}
\frac{d}{dt}(\rho\bm{u}) & =\bm{J}\times\bm{B}\label{eq:Step4-Eq1}\\
\frac{dE}{dt} & =\bm{u}\cdot(\bm{J}\times\bm{B})\label{eq:Step4-Eq2}
\end{align}
are solved by using a standard second-order ODE solver with two half-time
step updates that conserves the energy and momentum \cite{villa2011multiscale}.
\end{itemize}

\section{Results}

In this section, data from the improved, newly generated equation
of state is shown together with its implementation in the 1D and axisymmetric
plasma models.

\subsection*{EOS19 data}

\begin{figure}[tbh]
\includegraphics[width=1\columnwidth]{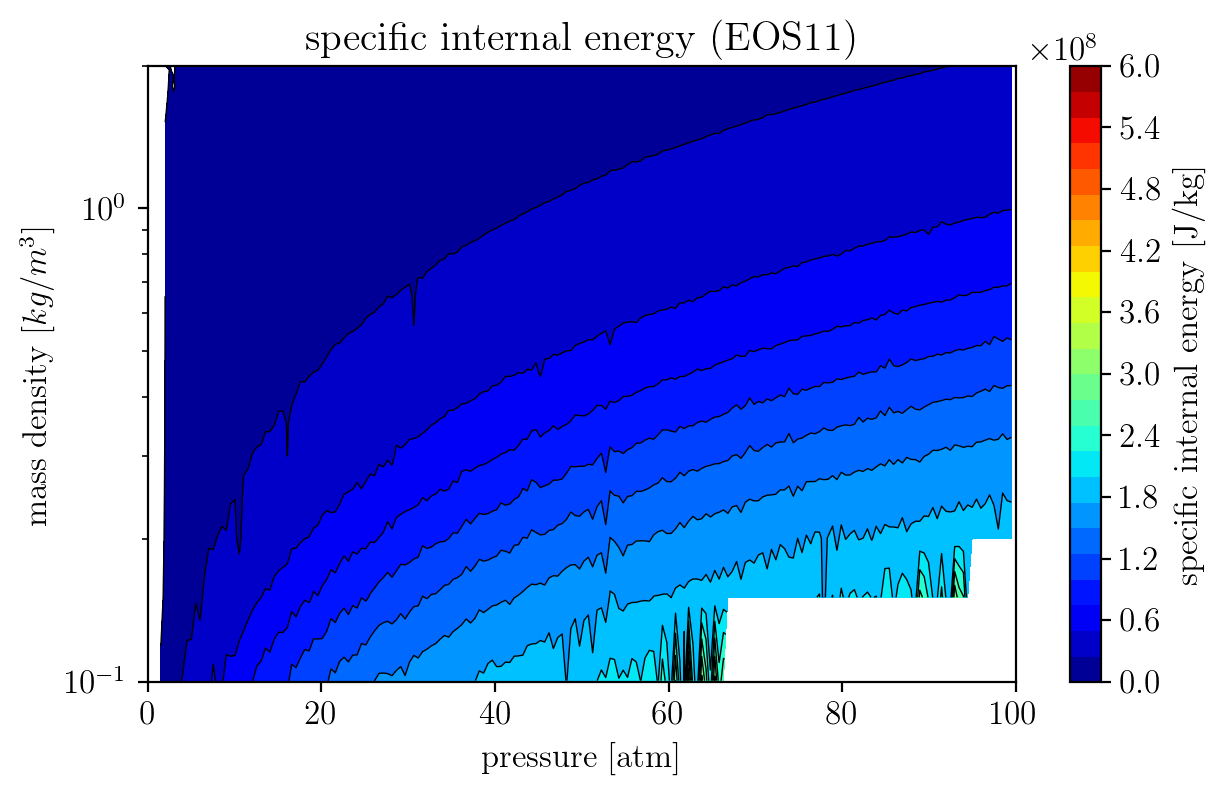}%

\includegraphics[width=1\columnwidth]{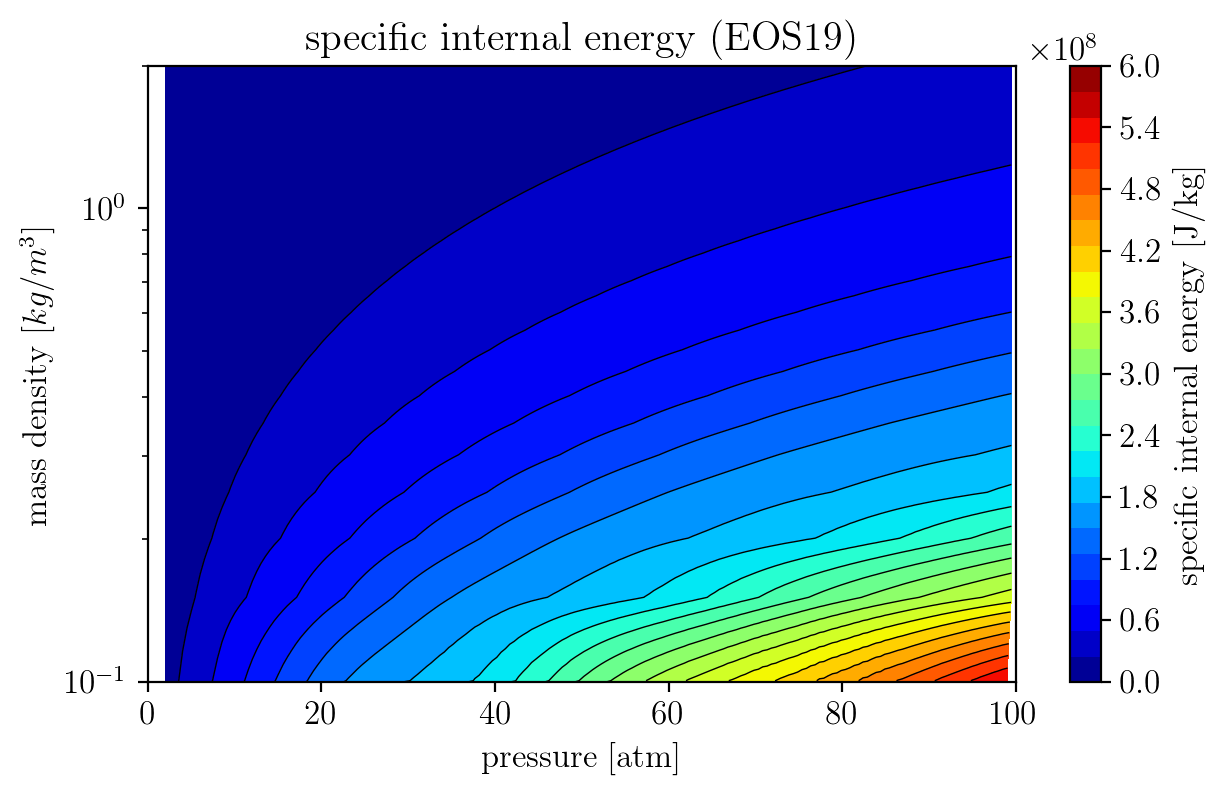}%
\caption{Heat map of the specific internal energy for EOS11 (top)
  and EOS19 (bottom) \label{fig:eos-comparison}}
\end{figure}

The specific internal energy generated by EOS19 for a given mass density
and pressure is shown in Fig. \ref{fig:eos-comparison}.
These values are compared to the publicly available data of the
equation of state developed by Villa, denoted as
EOS11~\cite{villa2011multiscale,VillaSourceCode}.  As there are no
classical phase transitions observed in air plasmas, a physically
accurate description requires smooth contour lines in any physical
variable.  Qualitatively, both equations of state show similar
behaviour with smooth contours in EOS19.  However, there are accuracy
issues in EOS11, which generate unphysical unsteady features at low
densities and high pressure as will be shown shortly.  The missing
region in EOS11 is due to a lack of sufficient data points for
reliable interpolation.

\subsection*{1D Model}

\begin{figure}[tbh]
\includegraphics[width=1\columnwidth]{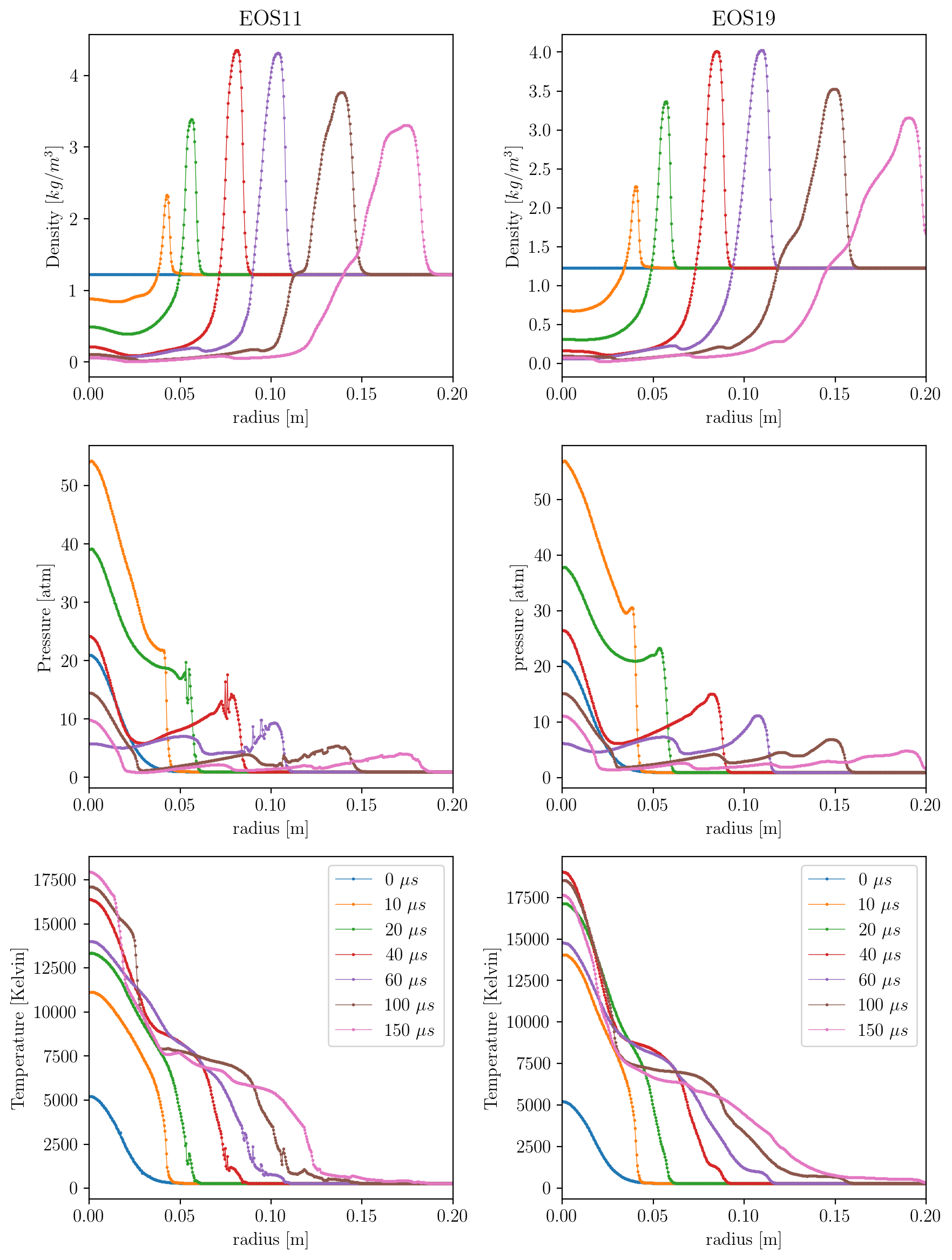}
\caption{Radial profiles $(N=500)$ for a component $V$ current for
  EOS11 (left) and the new EOS19 (right).  In both cases, the results
  qualitatively agree, however, the oscillations visiible
  at about 2,000\,K in the EOS11 results do not appear when using
  EOS19.\label{fig:Radial-profiles_1D_Evolution}}
\end{figure}

Simulations in 1D were calculated on a radial domain with $r\in[0,0.2\text{m}]$,
with initial data is chosen to represent a realistic air state with
mass density $\rho=1.225\text{kg/\ensuremath{\text{m}^{3}}}$, velocity
$u=0\text{m/s}$ and pressure $P(r)=P_{0}+P_{1}(r)$ with $P_{0}=101,325\text{Pa}$.
Initially, the centre of the domain is preheated, as a non-vanishing
electrical conductivity is needed to further heat the plasma sufficiently
via the Joule heating to solve \eqref{eq:currentPotentialConnection}.
This preheating condition is imposed by an added a Gaussian pressure
profile $P_{1}=2\times10^{6}\text{Pa}\cdot\text{exp}(-(r/r_{0})^{2})$
with $r_{0}=2\text{cm}$, which results in an initial Gaussian temperature
profile with $T_{peak}\approx5,000\text{K}$ in the centre of the
arc. Radial profiles with the oscillatory component $V$ current are
shown in Figure \ref{fig:Radial-profiles_1D_Evolution}
 for EOS19 and EOS11 in comparison. The mathematical form of a component
V current profile, and others applied in this work, are detailed
in the Table \ref{tab:Current-profiles-table}.
\begin{table}[tbh]
\begin{centering}
{\tiny{}}%
\begin{tabular}{|c|c|c|}
\hline 
{\tiny{}Component}  & {\tiny{}$D$} & {\tiny{}$V\,\text{(Villa)}$}\tabularnewline
\hline 
\hline 
{\tiny{}$I(t)$} & {{\tiny{}$I_{0}\big(e^{-\alpha t}-e^{-\beta t}\big)\big(1-e^{-\gamma t}\big)^{2}$}} & {\tiny{}$I_{0}e^{-\alpha t}sin(\beta t$)}\tabularnewline
\hline 
{\tiny{}$I_{0}$ {[}A{]}} & {\tiny{}$106,405$} & {\tiny{}$218,000$}\tabularnewline
\hline 
{\tiny{}$\alpha$ {[}$s^{-1}${]}} & {\tiny{}$22,708$} & {\tiny{}$4,137.95$}\tabularnewline
\hline 
{\tiny{}$\beta$ {[}$s^{-1}${]}} & {\tiny{}$1,294,530$} & {\tiny{}$114,866$}\tabularnewline
\hline 
{\tiny{}$\gamma$ {[}$s^{-1}${]}} & {\tiny{}$10,847,100$} & {\tiny{}-}\tabularnewline
\hline 
{\tiny{}time to peak {[}$\mu s${]}} & {\tiny{}$3.18$} & {\tiny{}$13.7$}\tabularnewline
\hline 
{\tiny{}time to half-peak {[}$\mu s${]}} & {\tiny{}$34.5$} & {\tiny{}$41.1$$(2^{nd}\text{peak})$}\tabularnewline
\hline 
\end{tabular}{\tiny\par}
\par\end{centering}
\caption{Current components applied in simulations lightning strike testing.
  Component $D$
  corresponds toa a single high-amplitude pulse with decreasing peak current
  as defined in \cite{arp54162005aircraft}.
  Additionally, a component $V$ current profile has been used in \cite{villa2011multiscale}
  and reflects a high-amplitude oscillatory exponentially decaying current
  profile.\label{tab:Current-profiles-table}}
\end{table}
Both models show similar behaviour, within the first
$10\text{\ensuremath{\mu}s}$, the plasma arc channel heats up quickly,
and the pressure increases rapidly reaching peak values of
$P_{peak}=55\text{atm}$ for EOS19 and almost $P_{peak}=60\text{atm}$
for EOS11. The formation of a radially expanding shock wave is
observed, which leaves a low-density region in the centre of the arc
channel. The speed of the shock wave is faster for EOS19, this
difference is likely to be associated with EOS11 showing the
incorrect internal energy at low temperatures. As the shock travels
outwards, it undergoes geometric expansion, hence reduces in
magnitude. At the centre of the arc, pressure and temperature show
pulsed behaviour with time, which is a result of the oscillatory
current profile, and this generates further pressure. The temperature
in the centre remains between $12,500\text{K}$ and $20,000\text{K}$
whilst current input remains high. EOS11 displays oscillatory spikes
in the pressure profile behind the leading shock wave; similar
oscillations can be seen in the corresponding temperature profiles.
These features occur in a temperature range between $T<1,000\text{K}$
and $T<3,500\text{K}$, corresponding to the region in which EOS11
shows an unphysical concentration of of ionised species. In contrast,
both temperature and pressure profiles for EOS19 are smooth everywhere
indicating an improvement to the accuracy of the results.  This
demonstrates the importance of an accurate representation of the
ionisation process in obtaining accurate behaviour within an air
plasma.

\begin{figure}[tbh]
\includegraphics[width=1\columnwidth]{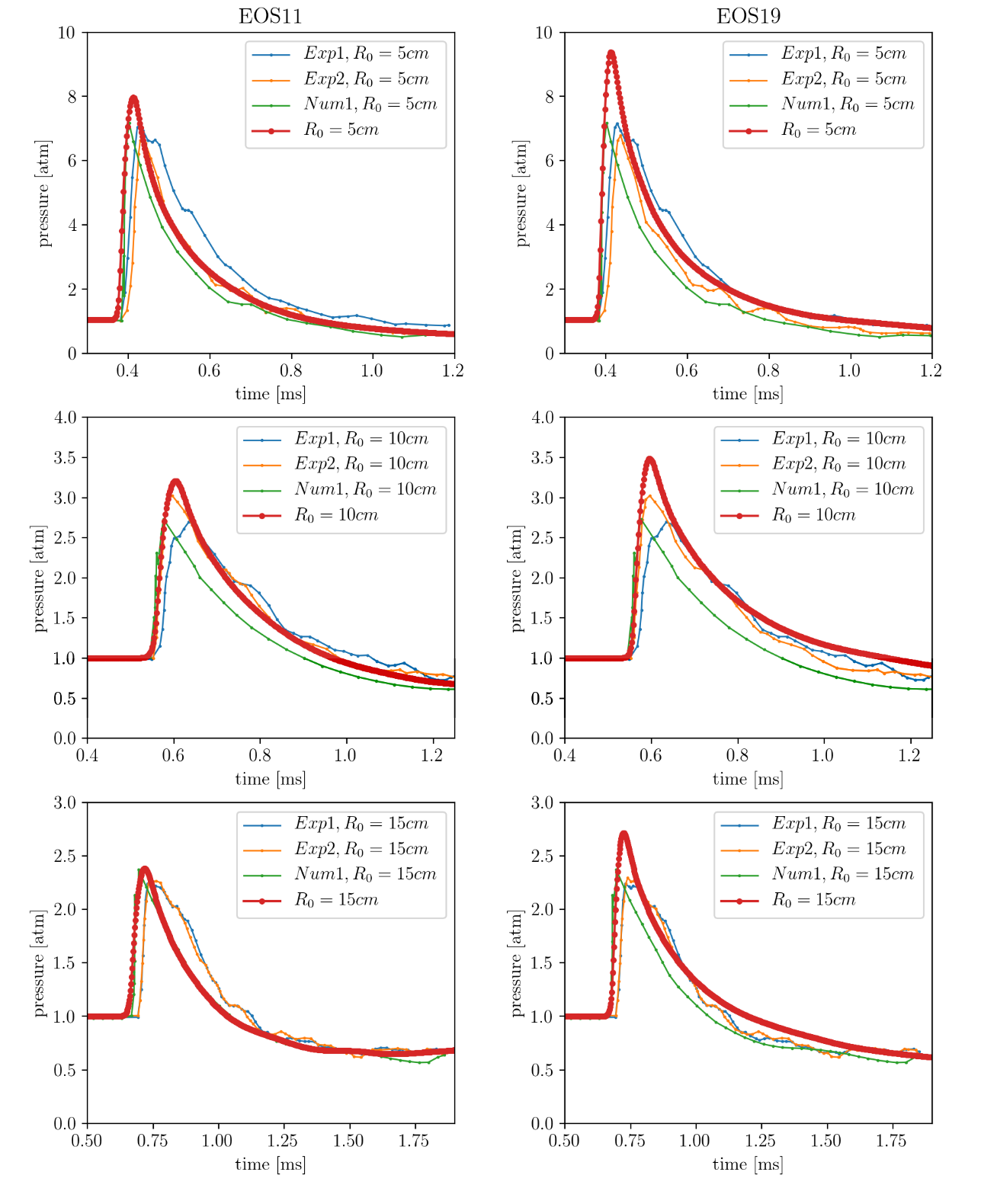}
\caption{Pressures calculated at the end of transducer tubes of length $L=15\text{cm}$
being $5\text{cm}$, $10\text{cm}$ and $15\text{cm}$ away from the
centre. Results in comparison for EOS11 (left) and EOS19 (right) with
experimental (`Exp1' and `Exp2') and numerical (`Num1') results from Villa et al.\ \cite{villa2011multiscale}.
\label{fig:Numerical-results-for_pressure-tubes-1D}}
\end{figure}

\noindent Validation of the model presented in
Section~\ref{subsec:Air-plasma-model} follows the approach of Villa et
al.~\cite{villa2011multiscale}, where pressure loading due to the
expansion of a cylindrical plasma arc is recorded. Specifically, in
their experimental set up, they drilled three holes into the substrate
at radial distances 5cm, 10cm and 15cm from the arc attachment
point. These holes were connected by a 25cm tube to pressure
transducers which recorded the effects of the passing shock wave and
subsequent post-shock behaviour. Villa et al.\ compared these
experimental results with their numerical simulations using a
one-dimensional approximation for the tube; here this test is used to
validate data for the equation of state and plasma model. Results for
the pressure at the three locations using the model presented here,
both with EOS11 and EOS19, are shown in Figure
\ref{fig:Numerical-results-for_pressure-tubes-1D}.
The two models show qualitatively similar behaviour, although EOS19
predicts a higher pressure peak than EOS11 upon arrival of the initial
wave. However, the subsequent pressure decrease after this peak is
then more consistent with experimental measurements from Villa
\cite{villa2011multiscale} for EOS19. The finite width of the tubes is
not modelled in the one-dimensional simulation of the behaviour, hence
it is likely that the experimental pressures near the peak are
slightly flattened due to geometric effects not captured by the
numerical models. 
\\

\subsection*{Multimaterial lightning code }

Having validated EOS19 in isolation, it is now used within the
multimaterial model introduced in Millmore and Nikiforakis \cite{millmore2019multiphysics}.
Using this, it is possible to investigate the effect of the EoS on
simulations of arc attachment, both qualitatively and quantitatively.

The experiments of Martins \cite{martins2016etude} provide imaging of
the arc evolution attaching to various substrates, as well as
measurements of the expansion of both the shock wave generated by the
attachment, and of the arc itself.  These tests use a component~$D$
current, as defined in table~\ref{tab:Current-profiles-table}, and a
simulation in line with a typical laboratory set up is considered, a
conical electrode geometry with a flattened tip of radius
$2.3\text{mm}$ is used.  The substrate (see setup in Figure
\ref{fig:Geometry__2D_plane}) has a thickness of $\delta=2\text{mm}$
and is modelled as an elastoplastic solid, as described in
Millmore and Nikiforakis \cite{millmore2019multiphysics}. 

For all tests, initial data in the air plasma, outside of the arc, is
$\rho=1.225\text{kg/\ensuremath{m^{3}}}$, $\bm{u}=\bm{0}\text{m/s}$
and $P_{0}=101,325\text{Pa}$. For the arc channel to be initially
conductive, the air is preheated to $8,063\text{K}$, within a thin
region of radius $r_{0}=2\text{mm}$, for the full height between the
electrode and the substrate. This value corresponds to
$P=40\times101,325\text{Pa}$ and $\sigma\approx312\text{Sm}^{-1}$.  The domain beneath the substrate is air,
modelled as an ideal gas with $\gamma=1.4$. 

Martins considers arc attachment to two materials, aluminium and CFRP;
here results are presented using the techniques described in
section~\ref{subsec:Mathematical-formulation-plasma-discharge}, using
a quasi-isotropic model for CFRP, which is henceforth referred to as a
low-conductivity substrate. 

\begin{figure}[tbh]
\includegraphics[width=1\columnwidth]{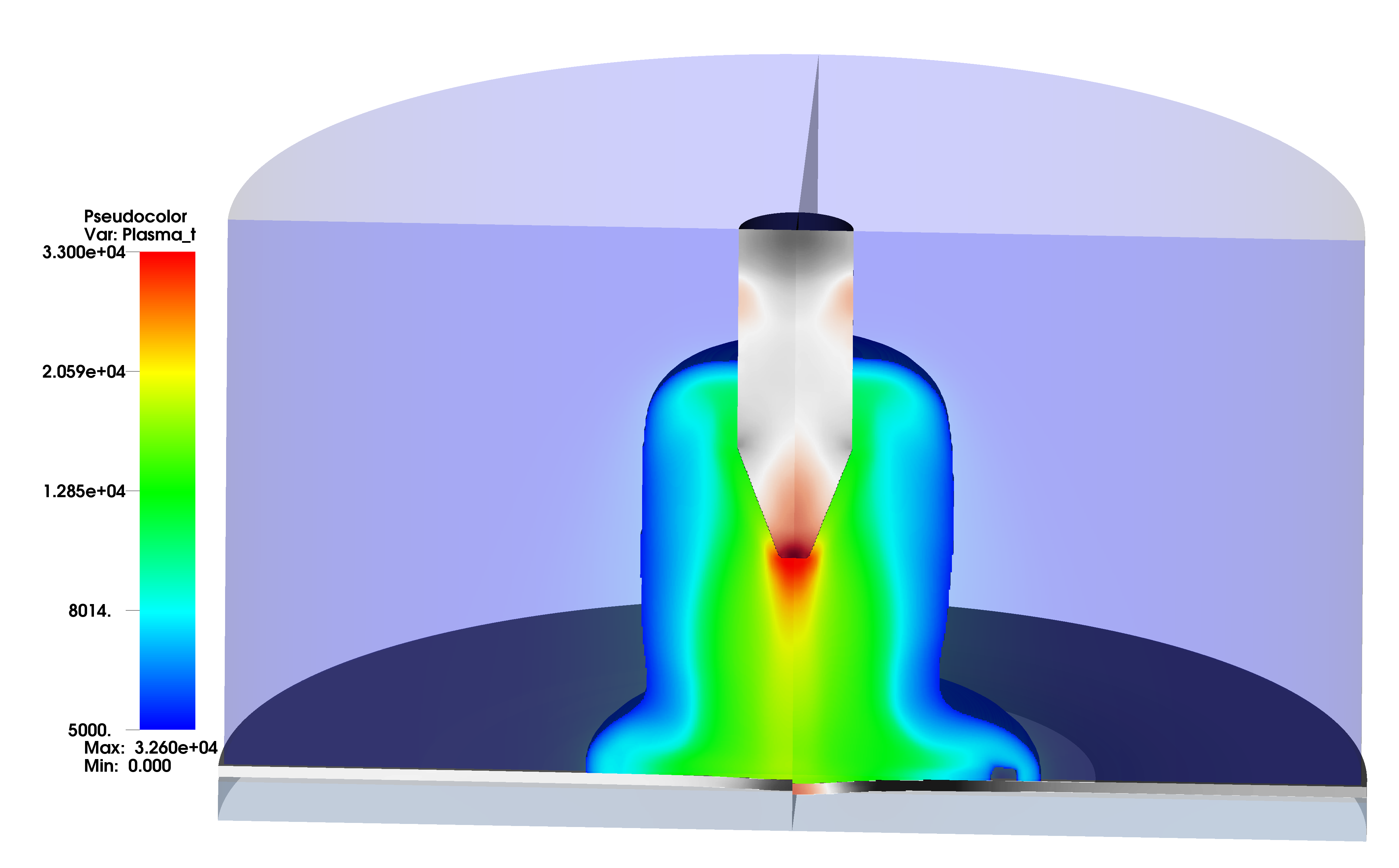}

\includegraphics[width=1\columnwidth]{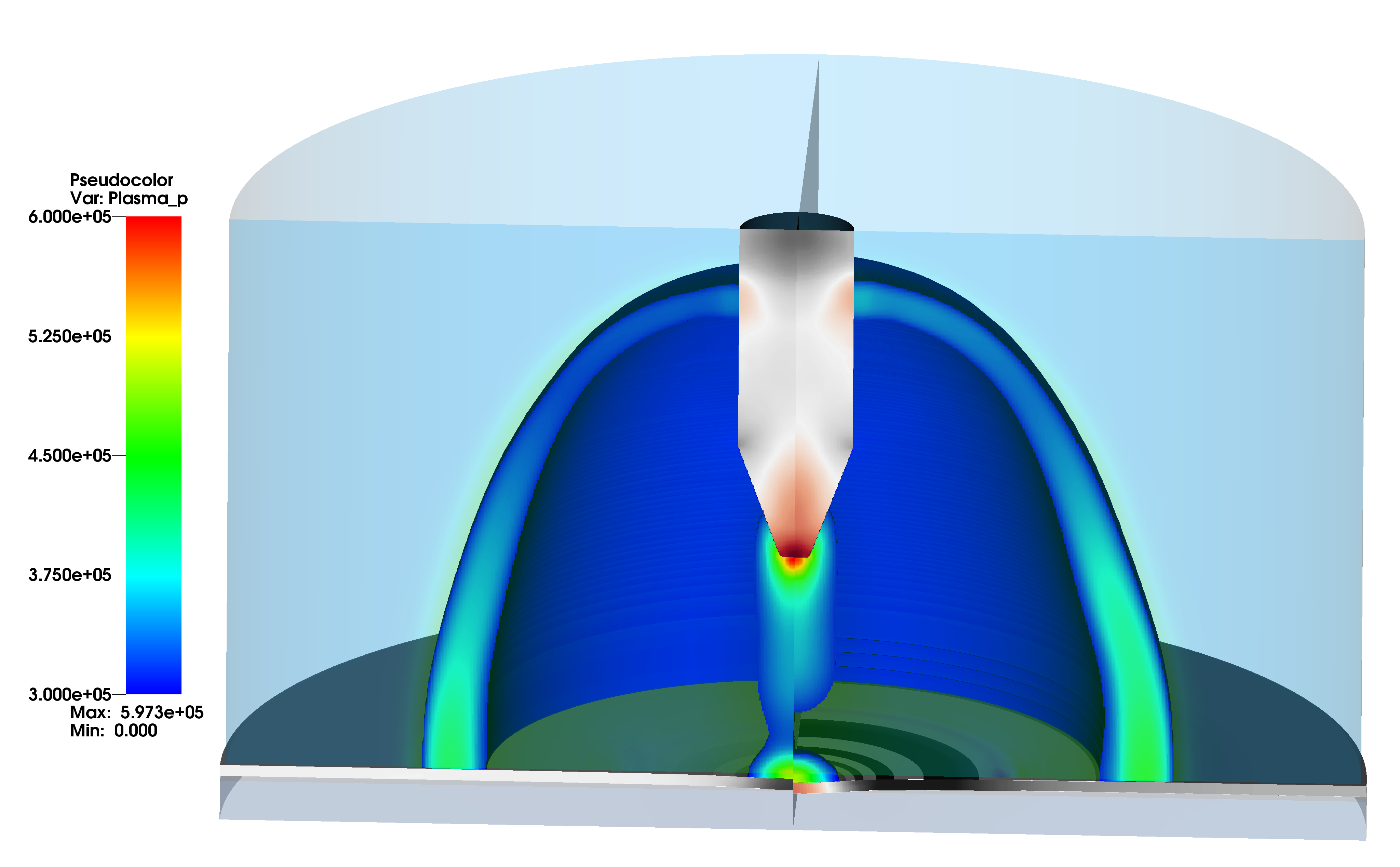}

\caption{Results for the plasma arc interacting with an aluminium panel (left
half) and CFC panel (right half) at
$t=50\text{\text{\ensuremath{\mu}s}}$.  The top plot shows
temperature and the bottom plot shows pressure.\label{fig:CombinedResults}}
\end{figure}

Figure
\ref{fig:CombinedResults}
shows illustrative temperature and pressure profiles for the arc attachment to
both substrates at $t=50\text{\ensuremath{\mu}s}$ using EOS19.  

\begin{figure}[tbh]
  \includegraphics[width=1\columnwidth]{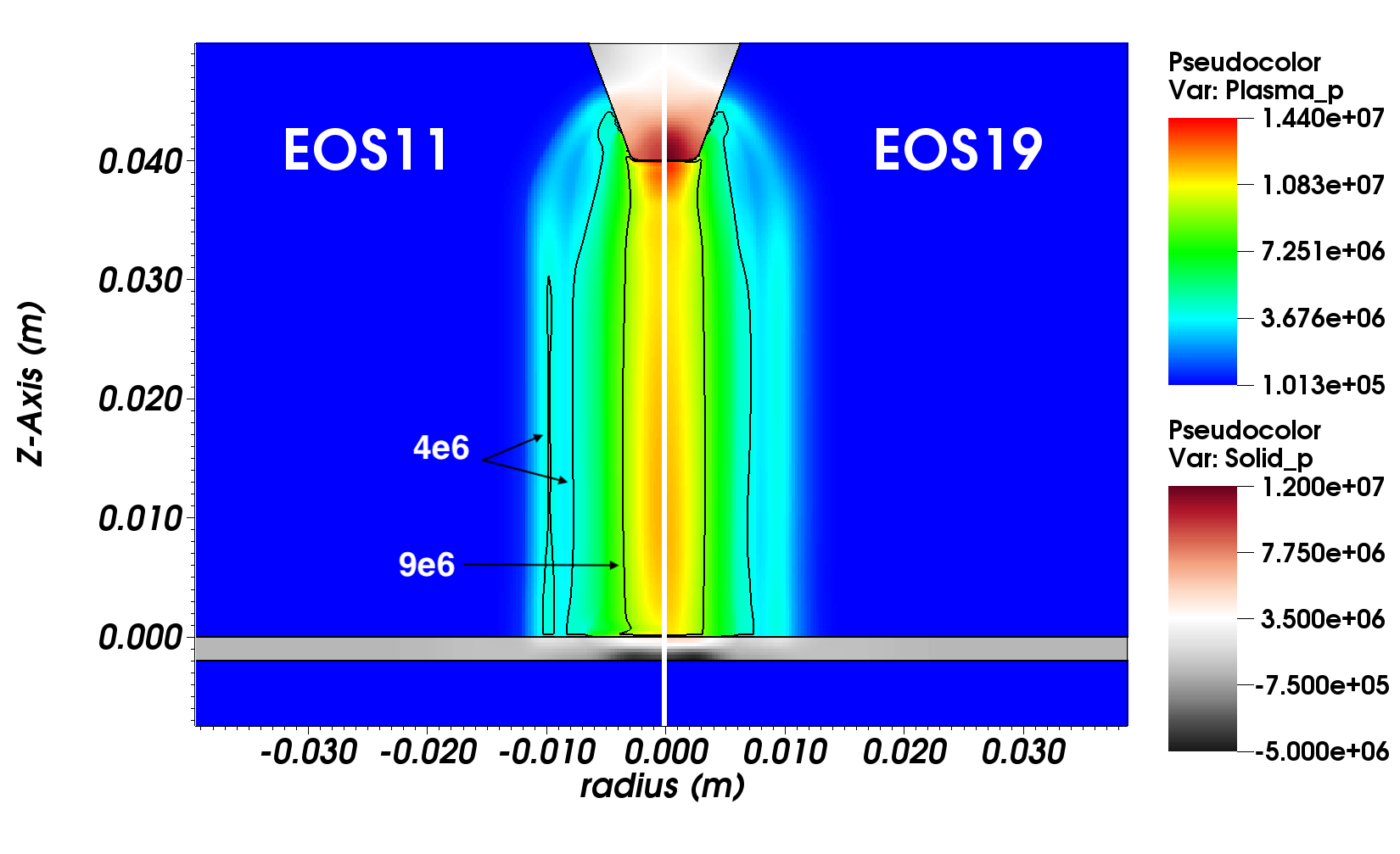}

  \includegraphics[width=1\columnwidth]{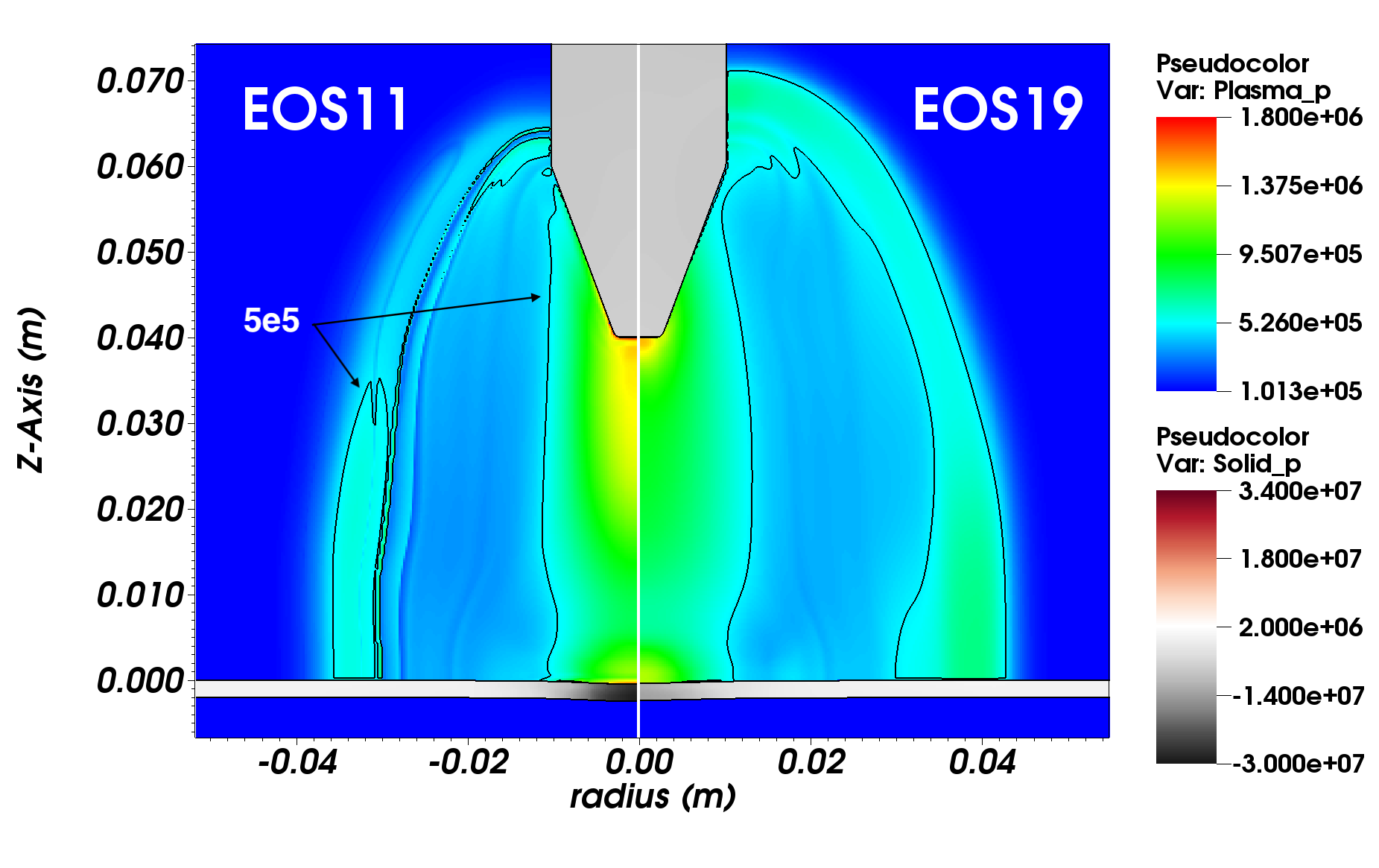}
  \caption{Pressure distribution and stresses for a component $D$ plasma arc
interacting with an aluminium panel.  The top image is plotted after
3.18$\mu$s, and the bottom plot after 25$\mu$s.\label{fig:Pressure-comparison-Component-D}}
\end{figure}

\begin{figure}[tbh]
  \includegraphics[width=1\columnwidth]{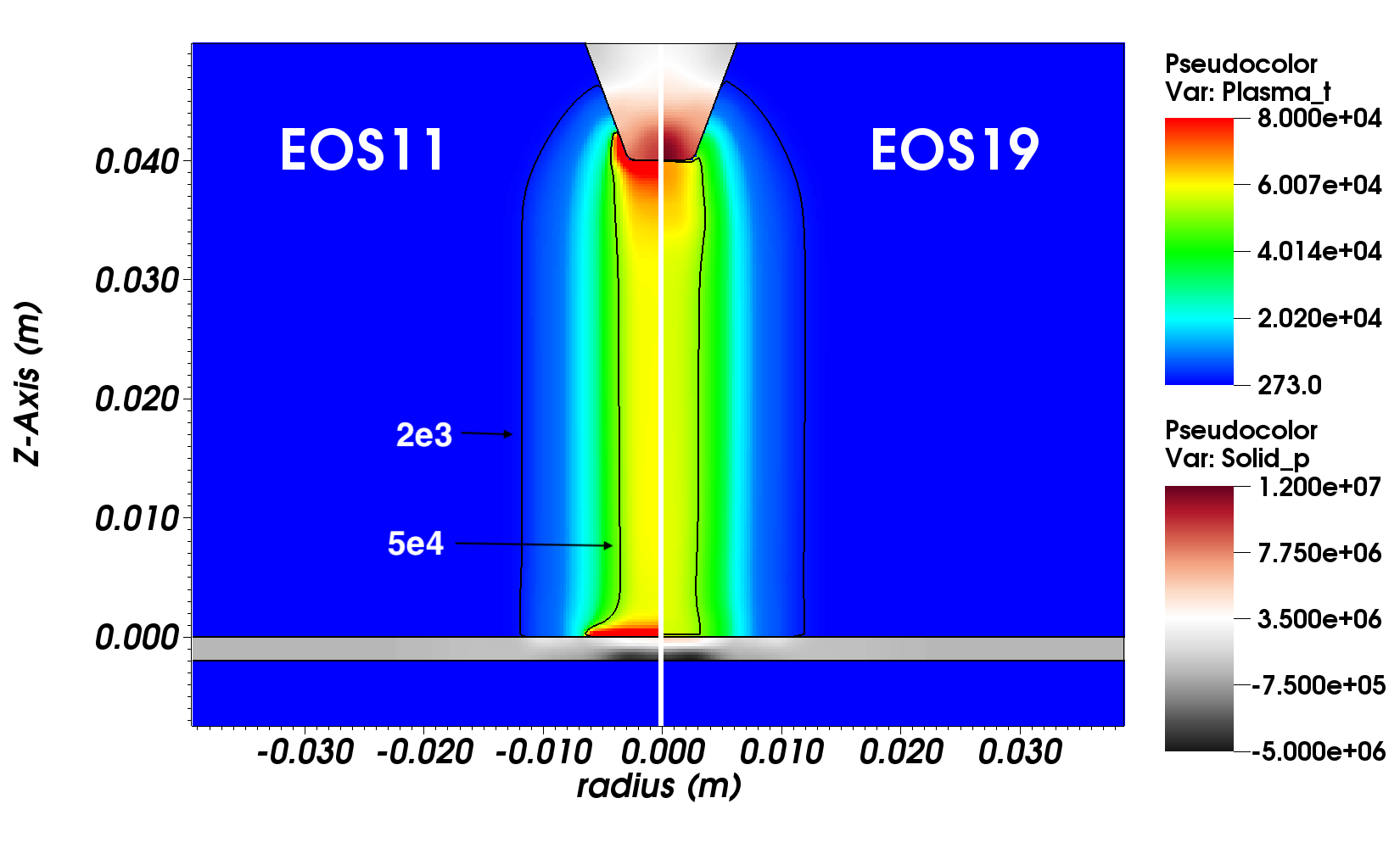}

  \includegraphics[width=1\columnwidth]{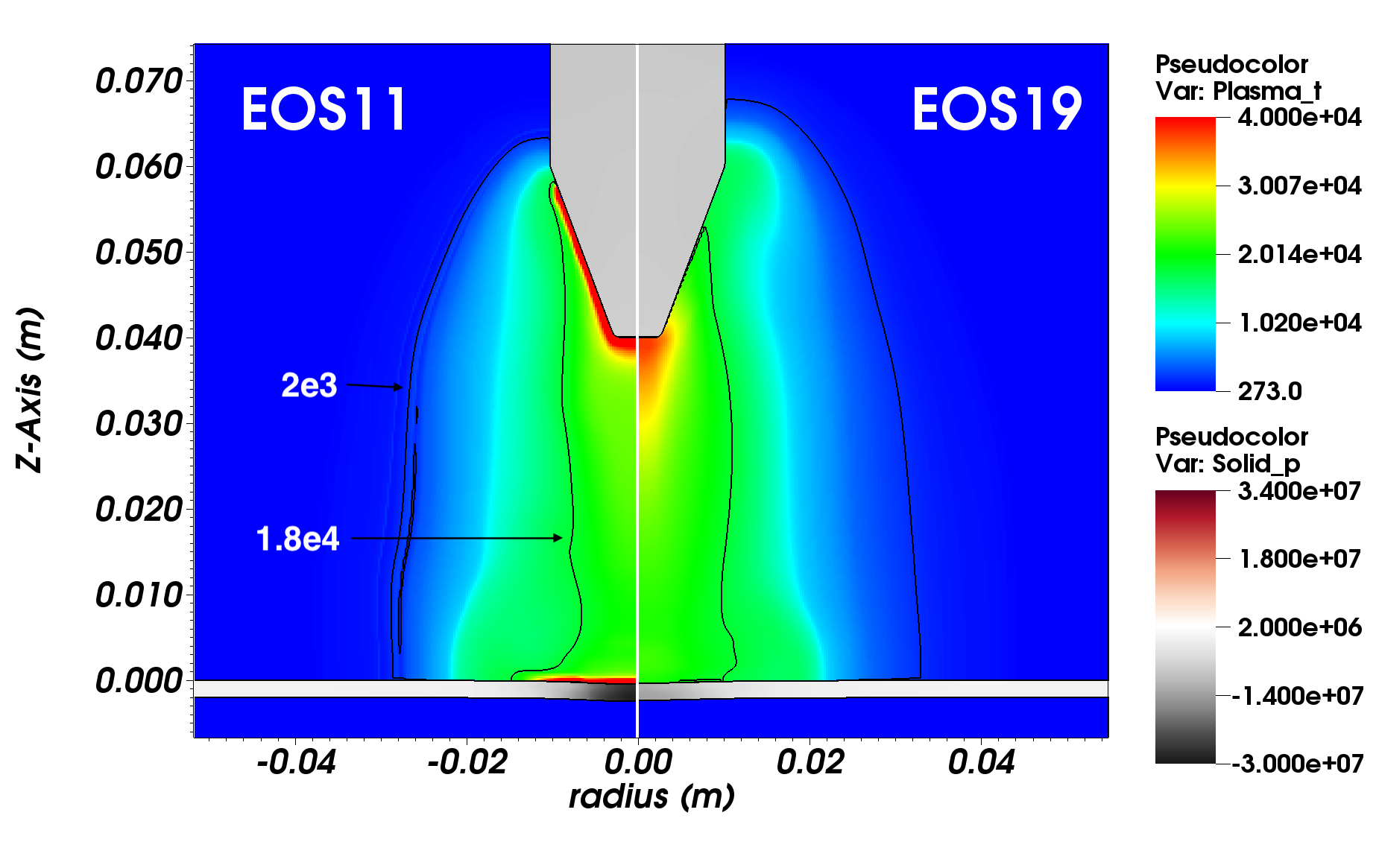}
  \caption{Temperature distribution and stresses for a component $D$ plasma
arc interacting with an aluminium panel. Temperatures at hot spots
in EOS11 are substantially higher than $80,000\text{K}$. The top image is plotted after
3.18$\mu$s, and the bottom plot after 25$\mu$s.\label{fig:Temperature-comparison-for-Component-D}}
\end{figure}

In order to understand the difference between the two equations of state
for arc attachment, the initial data described above is simulated with
both EOS11 and EOS19.  Results are shown for arc attachment to an
aluminium substrate at two time instances, $t = 3.18\text{\ensuremath{\mu}s}$
and $t = 25\text{\ensuremath{\mu}s}$, for pressure and temperature in Figures
\ref{fig:Pressure-comparison-Component-D} 
 and \ref{fig:Temperature-comparison-for-Component-D} respectively.
The first of the two times corresponds to the peak current input time
and there are similar pressure profiles for both EOS11 and EOS19.  It
is noted that EOS11 results in a lower pressure directly above the
substrate than for EOS19, whilst the pressure within the shock wave
appears to be higher, indicated by the contour plotted at
$P = 4\times10^6$Pa.  This higher pressure in the peak is likely a
result of the oscillatory behaviour within the shock wave shown in
Figure~\ref{fig:Radial-profiles_1D_Evolution}.  However, there are
clear differences in the temperature profiles between the two
equations of state, even at this early time.  EOS11 displays
unphysical hot spot-like domains, with temperatures of up to
$150,000\text{K}$, close to electrode and substrate, most likely due
to incorrect thermodynamic states obtained for coupling the materials
boundaries in EOS11.
Such hot spots are not observed for for EOS19, where temperatures do
not exceed $67,500\text{K}$, and nor are they evident in the
experimental work of Martins, where high radiation associated with
high temperature would be expected to be visible.  Additionally, EOS19
predicts a more uniformly cylindrical shape of the arc channel,
including at the attachment point, as is the expected for arc
attachment to aluminium \cite{tholin2015numerical}.

\noindent At $t=25\text{\text{\ensuremath{\mu}s}}$ the current flowing
through the arc has decreased substantially. A characteristic pressure
shock wave has emerged, which moves radially outwards, as can be seen
in Figure \ref{fig:Pressure-comparison-Component-D}.  At this time,
there are clear differences in the the shock front comparing EOS11 and
EOS19.  For EOS 11, the shock wave has not travelled as far
(approximately $r\approx4\text{cm}$, compared to
$r\approx4.5\text{cm}$ for EOS19), and the pressure within the wave is
lower.  The contour plotted at $P = 5\times10^5$Pa clearly shows the
unphysical oscillatory behaviour seen in the plasma discharge results
in Figure \ref{fig:Radial-profiles_1D_Evolution}.
As expected, EOS19 gives
smooth profiles without oscillatory regions throughout the shock
wave. 

There are similar differences in the temperature profile in Figure
\ref{fig:Temperature-comparison-for-Component-D}.  For EOS11, the
hot spots observed at earlier times are still present, and
have expanded in size.  These hot spots significantly alter the
behaviour of the attachment point to the aluminium substrate,
effectively widening the attachment region. 
For both equations of state, these results also show the aluminium
substrate mechanically deforming due to the pressure loading on
the panel.
\\

\begin{figure}[tbh]
\noindent\begin{minipage}[t]{1\columnwidth}%
\includegraphics[width=1\columnwidth]{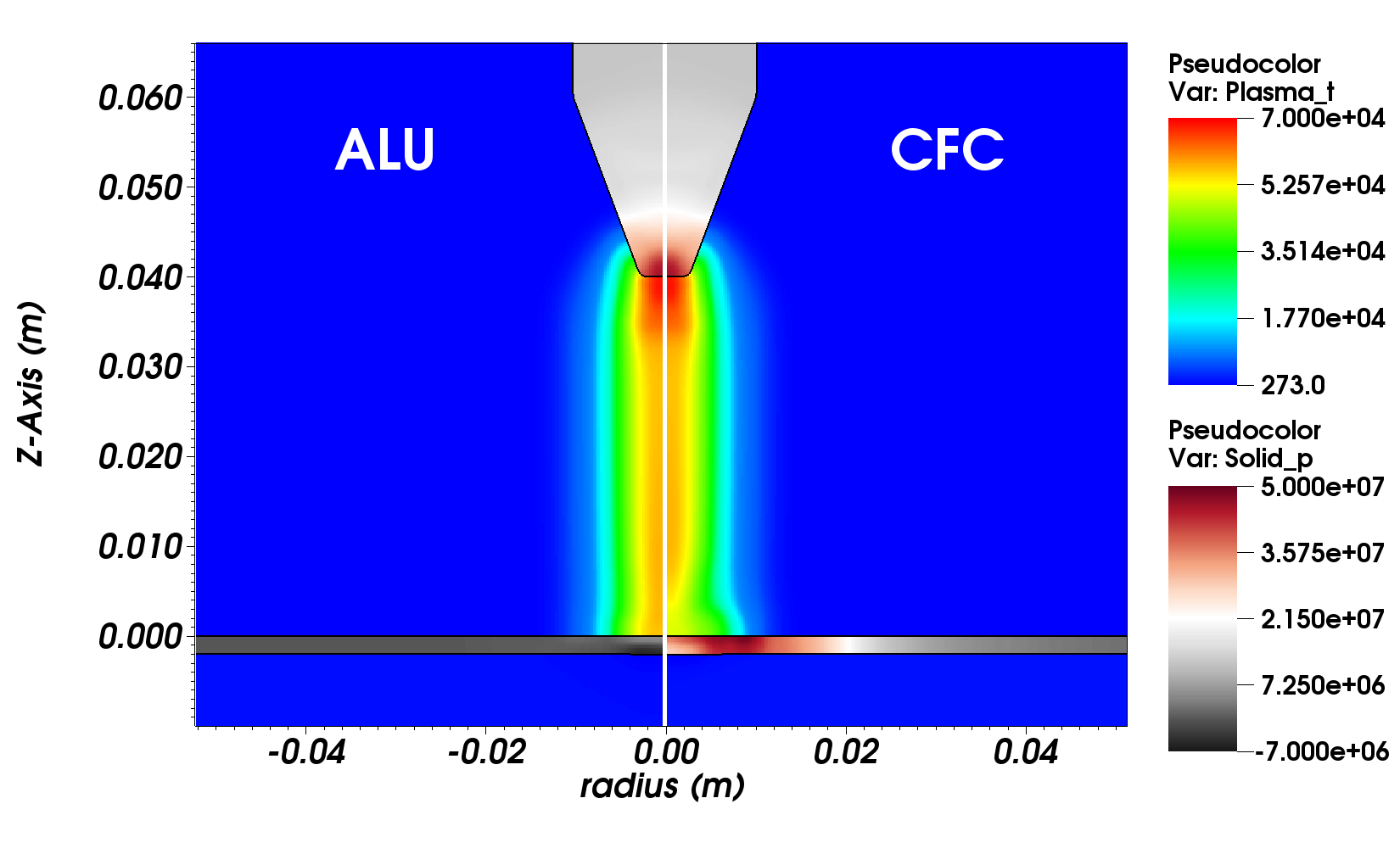}
\end{minipage}\medskip{}
\noindent\begin{minipage}[t]{1\columnwidth}%
\includegraphics[width=1\columnwidth]{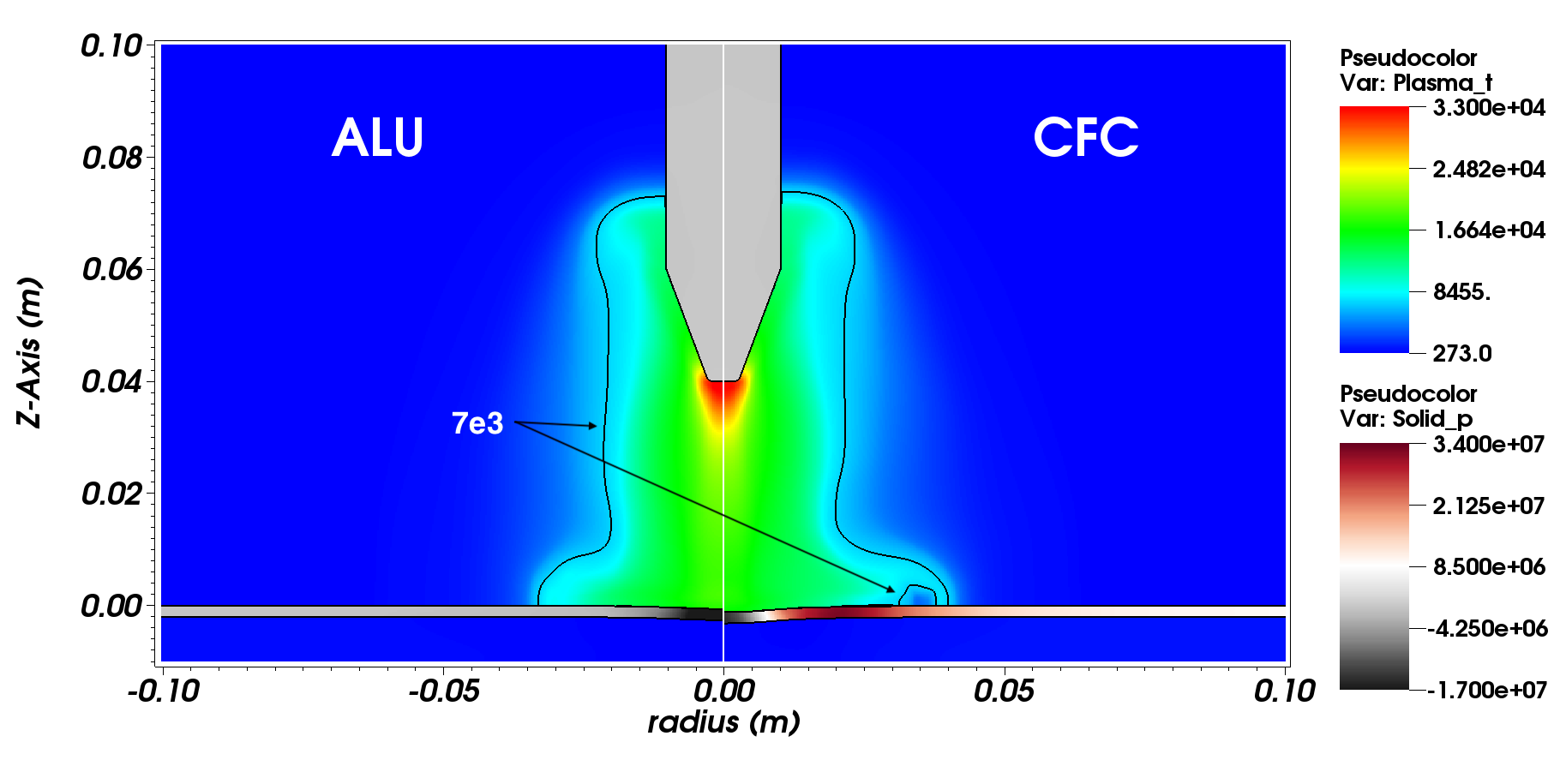}
\end{minipage}\caption{Temperature distribution and stresses for a component $D$ plasma
arc interacting with a carbon composite using EOS19.  The top image is plotted after
3.4$\mu$s, and the bottom plot after 46$\mu$s.\label{fig:CarbonTemperature}}
\end{figure}

Figure \ref{fig:CarbonTemperature}
\begin{figure}[tbh]
\noindent\begin{minipage}[t]{1\columnwidth}%
\includegraphics[width=1\columnwidth]{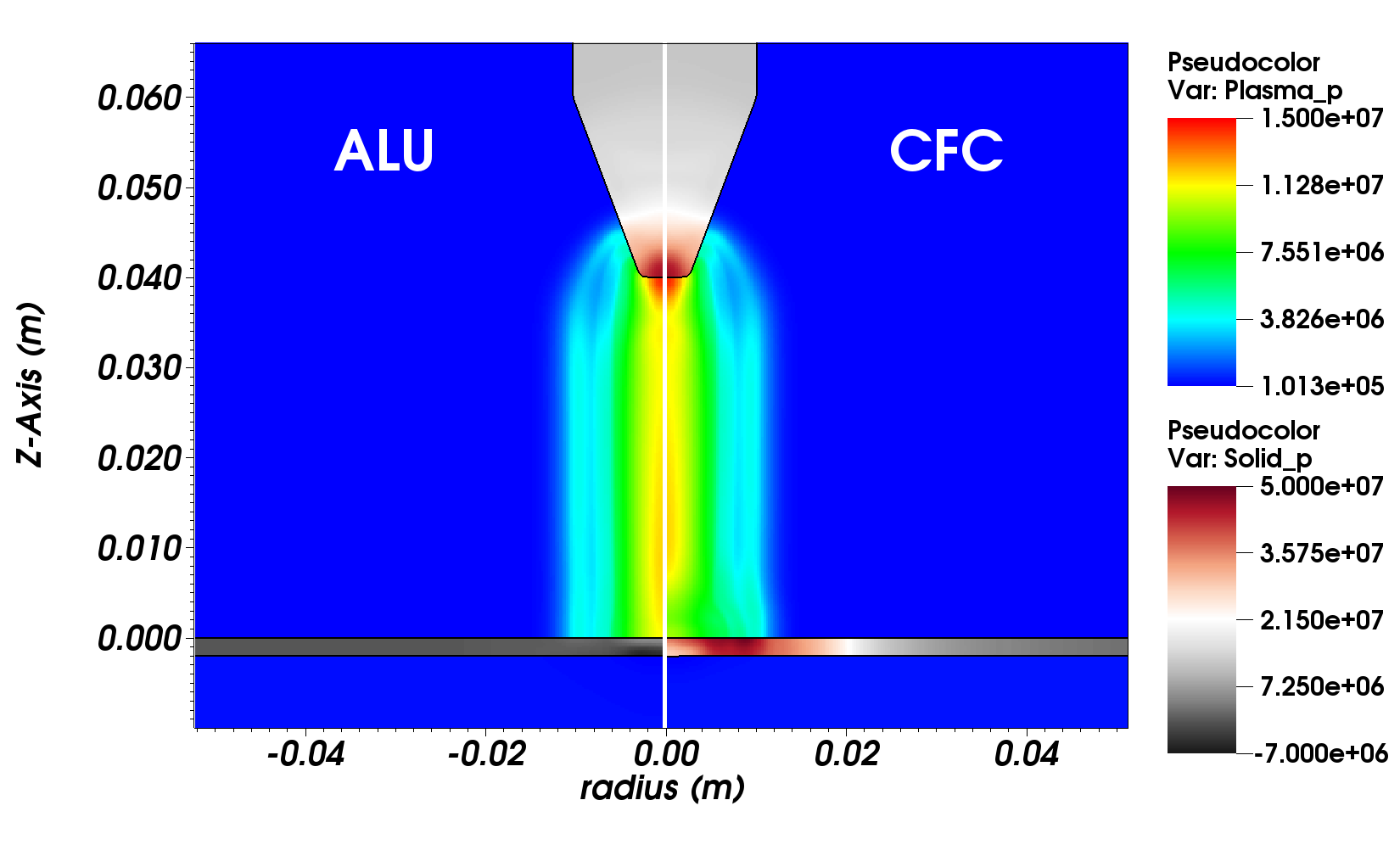}
\end{minipage}\medskip{}
\noindent\begin{minipage}[t]{1\columnwidth}%
\includegraphics[width=1\columnwidth]{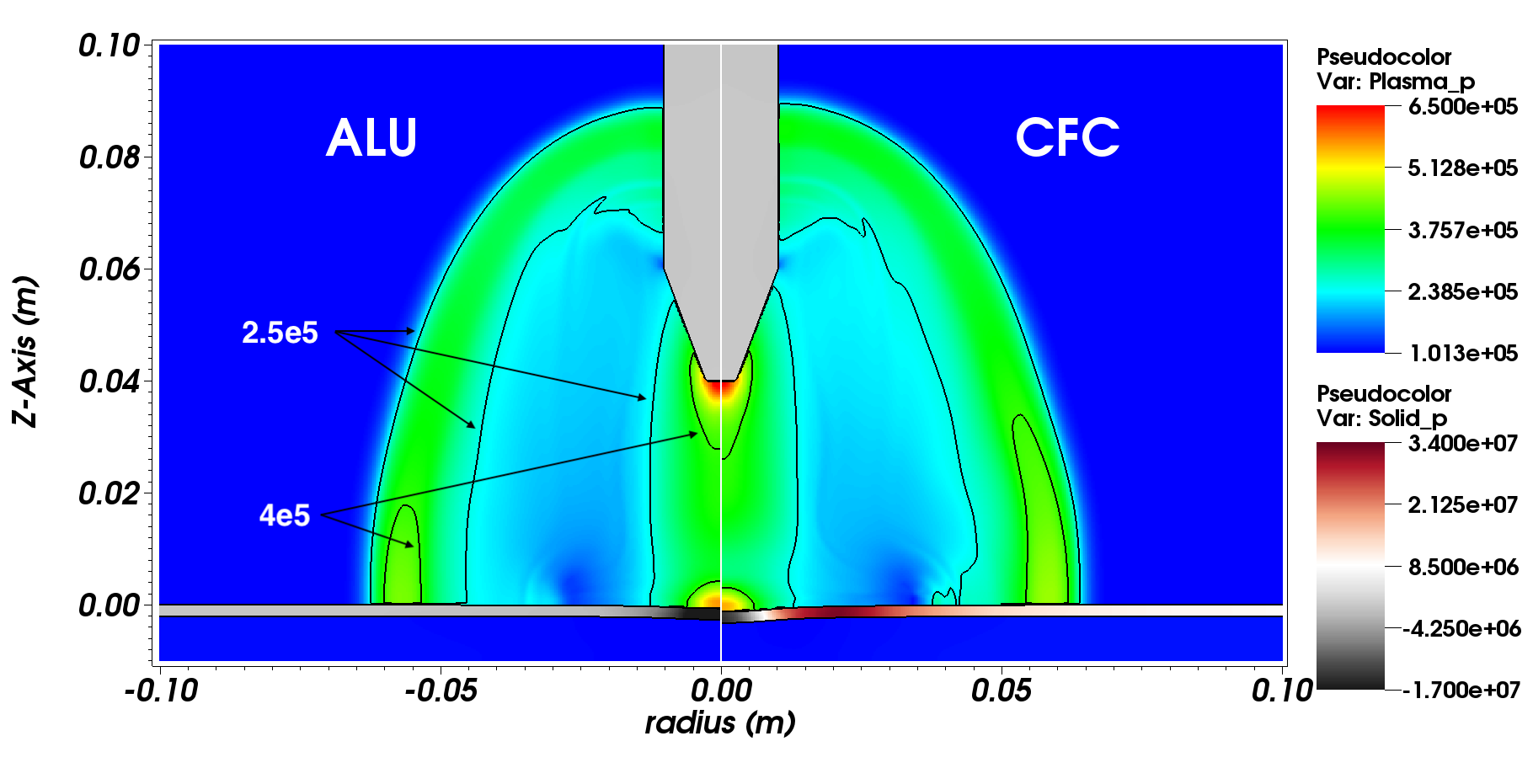}
\end{minipage}\caption{Pressure distribution and stresses for a component $D$ plasma arc
interacting with a carbon composite using EOS19.  The top image is plotted after
3.4$\mu$s, and the bottom plot after 46$\mu$s.\label{fig:CarbonPressure}}
\end{figure}
compares the temperature distribution for the two different substrate
materials; aluminium and the low-conductivity material. At
$t=3.4\text{\ensuremath{\mu}s}$, the peak temperature within the arc
is similar for both material, around
$70,000\text{K}$.  The arc shape does show small differences, however,
the root radius, where it attaches to the substrate, is broader for
the low-conductivity material, and this corresponds to a lower
temperature in the arc at this point.
Figure~\ref{fig:CarbonPressure} shows the pressure profiles at
corresponding times to the temperature plots in
Figure~\ref{fig:CarbonTemperature}.  The lower temperature at the
attachment point at $t=3.4\text{\ensuremath{\mu}s}$ corresponds
similarly to a lower pressure.  Away from the attachment point, the
arc profiles are similar for the two substrate materials.

Figures~\ref{fig:CarbonTemperature} and \ref{fig:CarbonPressure} also
show the pressure within the substrate; pressure rise is caused by a
direct loading from the arc, but also Joule heating within the
substrate.  It is clear that the magnitude of the pressure rise is
significantly higher for the low conductivity substrate, indicating
the vulnerability to greater damage displayed by such materials. 

After $t=46\text{\ensuremath{\mu}s}$, the pressure shown in
Figure~\ref{fig:CarbonPressure} shows that the radially expanding
shock wave is comparable in structure for both substrates.  However,
as with the earlier time, at the centre of the arc, pressure is lower
directly above the substrate.  The temperature plot in
Figure~\ref{fig:CarbonTemperature} shows a significant difference in
the arc root structure between the two substrates.  This shows that
for the low conductivity substrate, the arc root is significantly
wider at the attachment point.  This qualitatively distinct behaviour
of the plasma arc root channel between the two substrates is observed
experimentally~\cite{martins2016etude, tholin2015numerical}.  Of
particular interest for the attachment to the low conductivity
substrate is the feature which arises in arc root, close to the panel,
particularly visible in the temperature profile shown in
Figure~\ref{fig:CarbonTemperature}; an emerging filament, which
separates from the primary arc channel. The physical mechanism behind
this is that the plasma arc provides a favourable path with lower
electrical resistance opposed to the direct path to the
substrate. Such a filament is a characteristic phenomenon for carbon
composite panels, which has been observed in experiments by Martins
\cite{martins2016etude}, where the evolution of the visible arc root
radius was measured over time for both substrate materials.

The large pressure rise in the low conductivity material is again
visible at after $t=46\text{\ensuremath{\mu}s}$, and due to the wider
arc in this case, the extent of the high pressure region is wider in
comparison to the aluminium substrate.  Additionally, throughout the
evolution, the lower density of the low conductivity material can be
seen to result in greater deformation of the panel.
\\ 

\begin{figure}[tbh]
  \noindent\begin{minipage}[t]{1\columnwidth}%
    \begin{center}
        \includegraphics[width=0.6\columnwidth]{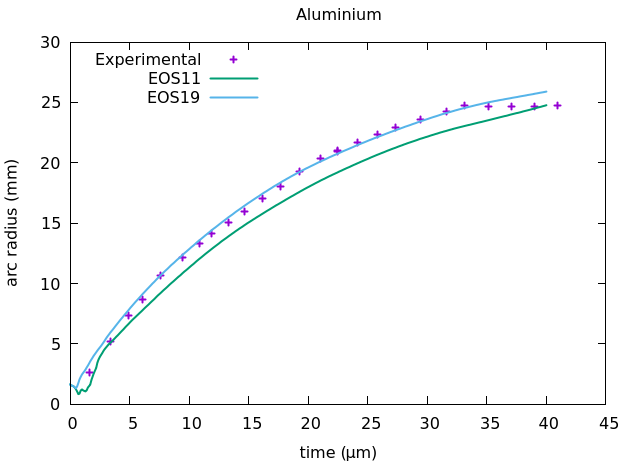}
    \end{center}
  \end{minipage}\medskip{}
  \noindent\begin{minipage}[t]{1\columnwidth}%
    \begin{center}
        \includegraphics[width=0.6\columnwidth]{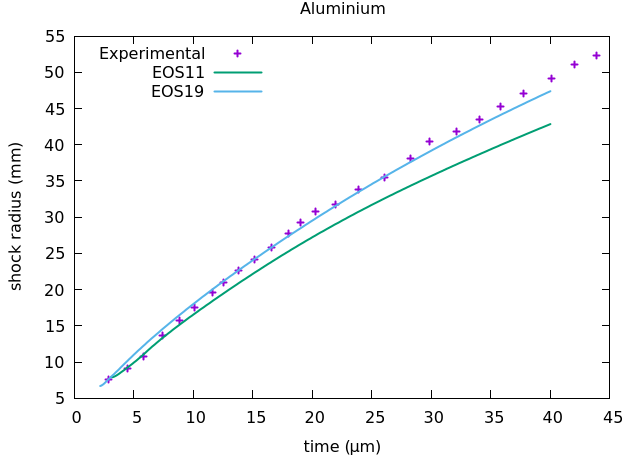}
    \end{center}
  \end{minipage}\caption{Comparison of the arc evolution for
    attachment on aluminium for EOS11 and EOS19.  It is clear that
    EOS19 captures the evolution behaviour well, with the faster
    expansion of the shock wave demonstrated in Figure
    \ref{fig:Pressure-comparison-Component-D} matching experimental
    values.  The top plot shows the expansion of the arc radius,
    whilst the bottom plot shows the shock radius.}
  \label{fig:radiusalu}
\end{figure}
In addition to the qualitative improvement shown by EOS19 for arc
attachment simulations, a comparison is made to the measurements of
Martins~\cite{martins2016etude}.  The expansion of both the arc and
the shock wave was measured through optical detection of the
attachment process.  Figure~\ref{fig:radiusalu} compares these
quantities for an aluminium substrate; it is clear that both equations
of state follow similar patterns, with EOS19 demonstrating an
improvement in accuracy.
\begin{figure}[tbh]
  \noindent\begin{minipage}[t]{1\columnwidth}%
    \begin{center}
        \includegraphics[width=0.6\columnwidth]{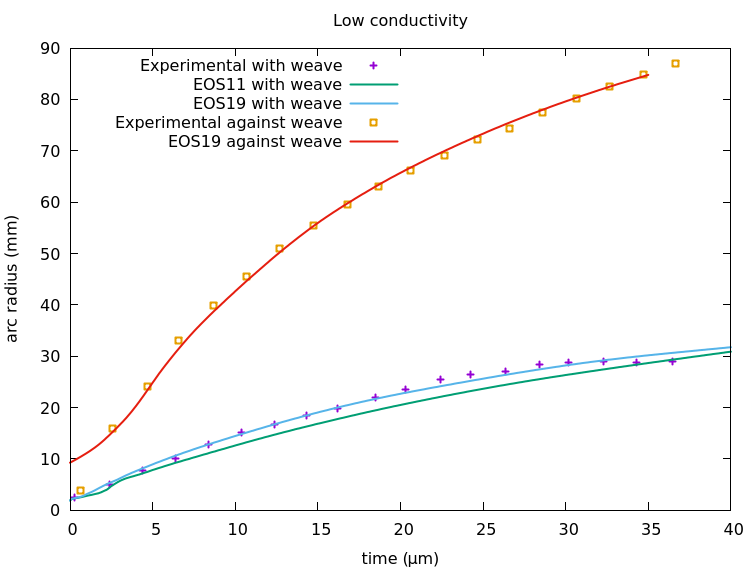}
    \end{center}
  \end{minipage}\medskip{}
  \noindent\begin{minipage}[t]{1\columnwidth}%
    \begin{center}
        \includegraphics[width=0.6\columnwidth]{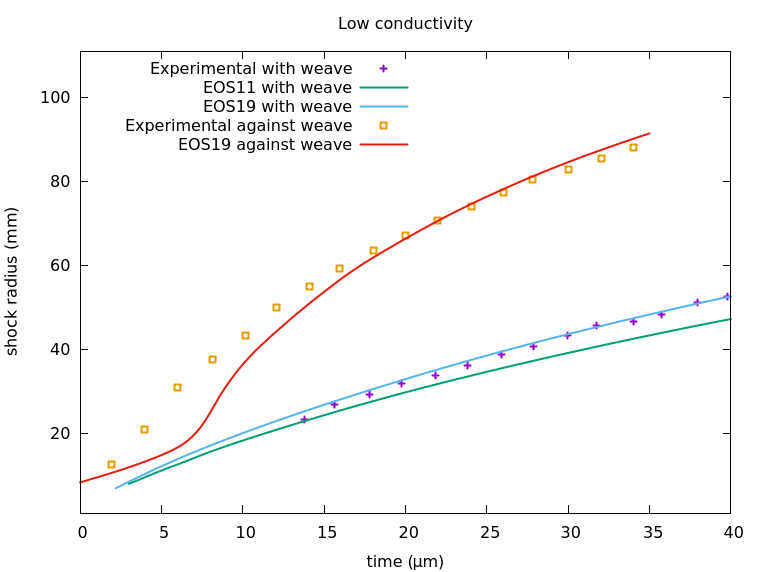}
    \end{center}
  \end{minipage}\caption{Comparison of the arc evolution for
    attachment on the low conductivity substrate for EOS11 and EOS19.
    Results are shown for two cases, with electrical conductivity
    corresponding to parallel and perpendicular alignment with a
    composite weave.  In the parallel `with weave' case, the
    improvement shown by EOS19 is clear.  In the `against weave' case,
    EOS11 was unable to simulate this case, whilst EOS19 again shows
    good agreement.  The top plot shows the expansion of the arc
    radius, whilst the bottom plot shows the shock radius.}
  \label{fig:radiuscomposite}
\end{figure}
In order to compare the arc and shock expansion to a low conductivity
substrate, following Millmore and
Nikiforakis~\cite{millmore2019multiphysics}, two measurements are
possible; arc growth parallel and perpendicular to the composite
weave.  The symmetry in these cases makes the cylindrically symmetric
model presented here valid.  The electrical conductivity used for this
comparison is $3 \times 10^4$S/m parallel to the weave and $100$S/m
perpendicular to it.  Results for these tests are shown in
Figure~\ref{fig:radiuscomposite}, where it is noted that in the case
of low `against weave' electrical conductivity in the substrate, EOS11
failed to produce a stable simulation.  For the parallel `with weave'
case, results are similar to those in Figure~\ref{fig:radiusalu}, with
EOS19 demonstrating an improved fit to the experimental results over
EOS11.  It is noted that the underprediction in shock radius at early
times in the `against weave' case is due to both the arc itself and
the shock being close together, hence it is difficult to measure an
accurate radius.  At later times, when the two features can be more
clearly distinguished, the simulation data matches the experiment well.

\section{Conclusions}

The primary objective of this work was to implement an improved
equation of state for air plasma, which can be used for various
applications involving the partial ionisation of air. In this paper,
emphasis was put on the computational multiphysics modelling of
solid-plasma interactions. Such numerical tools are of high interest
for aeronautics to study lightning strike on new aircraft
materials. The equation of state (EOS19) based on the theoretical
results of D'Angola et al.\
\cite{d2008thermodynamic,d2011thermodynamic}, who considered a
19-species air plasma. Their model is based on the Debye-Hückel theory
for (partially) ionised gas mixtures, and transport coefficients have
been calculated using the Chapman-Enskog theory.  Comparison to
existing open-source EoSs, e.g.\ Villa et al
\cite{villa2011multiscale}, show EOS19 can accurately capture the
evolution of a plasma arc in a non-oscillatory manner. Profiles of
various air plasma properties are shown to be accurate over a broad
temperature and pressure range. Additionally, the transport
coefficients for thermal conductivity have been included in the
database for usage in other models in the future.

\noindent To validate EOS19, an axisymmetric plasma discharge code
was used, simulating a plasma arc channel under laboratory conditions.
Results are presented in 1D, based on a model with fixed current
density from Villa et al.\ \cite{villa2011multiscale} and a more
realistic model in 2D axisymmetry, which generates the current and
magnetic field self-consistently based on the plasma evolution. The
results have been validated against experimental and numerical results
of others
\cite{villa2011multiscale,martins2016etude,plooster1971numerical,plooster1971numerical2,tholin2015numerical}.
It was additionally identified that small errors in the levels of
ionisation in an equation of state can lead to oscillatory behaviour;
the accuracy of EOS19 is demonstrated through the smooth profiles
which arise across the temperature and pressure range.

\noindent To further test EOS19, it was also applied within a state-of-the-art
multimaterial lightning simulation code, which additionally computes
the thermomechanical coupling between the plasma arc root and solid
aeronautical skin layers. Results are computed for an aluminium and
quasi-isotropic CFC panel under extreme lightning testing conditions
by inducing a standardised arc current that is used in lightning-protection
testing in laboratories. Results for both materials indicate good
agreement with experimental measurements by Martins \cite{martins2016etude},
and no unphysical high-temperature hot spots close to interfaces are
observed when applying EOS19.  Additionally, a quantitative
investigation of the expansion of both the shock wave caused by
attachment, and the arc itself, show that EOS19 achieves improved
accuracy on EOS11, matching the measured behaviour of these features well.

\noindent Future work will focus on improved radiative source term
within the arc in a computationally efficient manner. Additionally,
one may include possible effects associated with thermal conductivity.
Finally, the implementation of a fully 3D anisotropic equation of
state for carbon composite materials will be of high interest for
the future development of very complex materials in aeronautics.

\noindent To the best of the authors' knowledge, this state-of-the-art
lightning code in combination with EOS19 is the only multiphysics
code which considers aeronautical materials and
their coupled thermomechanical interaction with a
plasma arc channel within a single simulation. Hence, the
multimaterial lightning code with EOS19 functions as a reliable
computational tool to support the design and validation of
aeronautical materials concerning lightning protection.

\begin{acknowledgments}
  The authors would also like to thank Micah Goldade of BR\&T, and
  Carmen Guerra-Garcia of BR\&T and the Massachusetts Institute of
  Technology, for their input in improving the plasma model.
\end{acknowledgments}

\bibliography{ReferencesMPhil}

\end{document}